\documentclass[aps,pre,twocolumn,superscriptaddress,floatfix,showpacs]{revtex4}

\usepackage{amssymb,amsmath,graphicx,epstopdf,color}

\begin{document}

\title{Delayed feedback control of unstable steady states with high-frequency modulation of the delay}

\author{Aleksandar Gjurchinovski}

\email{agjurcin@pmf.ukim.mk}

\affiliation{Institute of Physics, Faculty of Natural Sciences and Mathematics, Sts.\ Cyril and Methodius University,
P.\ O.\ Box 162, 1000 Skopje, Macedonia}

\author{Thomas J\"ungling}

\affiliation{Institute for Cross-Disciplinary Physics and Complex Systems, IFISC (UIB-CSIC), University of the Balearic Islands, 07122 Palma de Mallorca, Spain}

\author{Viktor Urumov}

\affiliation{Partenij Zografski 46, 1000 Skopje, Macedonia}

\author{Eckehard Sch\"oll}

\email{schoell@physik.tu-berlin.de}

\affiliation{Institut f\"ur Theoretische Physik, Technische Universit\"at Berlin, 10623 Berlin, Germany}


\date{\today}

\begin{abstract}
We analyze the stabilization of unstable steady states by delayed feedback control with a periodic time-varying delay
in the regime of a high-frequency modulation of the delay. The average effect of the delayed feedback term in the control force 
is equivalent to a distributed delay in the interval of the modulation, and the obtained distribution depends on the 
type of the modulation. In our analysis we use a simple generic normal form of an unstable focus, and 
investigate the effects of phase-dependent coupling and the influence of the control loop latency on the controllability.
In addition, we have explored the influence of the modulation of the delays in multiple delay feedback schemes
consisting of two independent delay lines of Pyragas type.
A main advantage of the variable delay is the considerably larger domain of stabilization in parameter space.
\end{abstract}

\maketitle

\section{Introduction}

The possibility to stabilize unstable periodic or stationary states embedded in chaotic attractors has been 
elaborated more than two decades ago \cite{OGY90}. The main idea consists of using small external 
perturbations to force the system to follow one of its stable manifolds. These perturbations are applied 
at specific instances when the chaotic trajectory is close to the desired periodic orbit.
This seemingly straightforward theoretical concept has caused a revolution in applied nonlinear science. 
It has been realized that the control of chaos could have a significant 
outcome in real-world experiments, where one could generate different kinds of ordered behavior 
from an utterly erratic one \cite{SCH08}. 

A conceptually simple method to stabilize unstable equilibria and periodic 
orbits is the time-delayed feedback control (TDFC) introduced by Pyragas \cite{PYR92, PYR06}. Here, the 
perturbation has the form of a continuous feedback constructed from the difference 
between some suitable scalar signal obtained from the system and the same signal delayed by a constant time 
$\tau$. The difference signal is amplified by the gain factor $K$ and then re-injected 
into the original system. For a certain choice of the feedback gain $K$ and the delay time $\tau$, the control 
of the unstable state can be realized, in which case the feedback force vanishes by construction, making the 
method essentially noninvasive.

Since a detailed knowledge of the target state is not required and the controller is very 
robust with respect to noise, the Pyragas control has become one of the most popular control methods used in experiments and even technological applications.
The application is quite diverse, and includes stabilization of unstable states in 
electronic chaotic oscillators \cite{PYR93,CEL94,GAU94}, 
mechanical pendulums \cite{CHR97,PAU09},
laser systems \cite{BIE94,DAH10,SCHI11}, 
electrochemical systems \cite{PAR99,ZHA08}, 
drift waves in a magnetized laboratory plasma \cite{GRA00}, 
chaotic Taylor-Couette flow \cite{LUT01}, 
cardiac systems \cite{HAL97,BER07}, 
ferromagnetic resonance systems \cite{BEN02},
gas discharge systems \cite{PIE96,WEI04}, 
controlling helicopter rotor blades \cite{KRO00},
controlling the walking mechanism of a robot \cite{SUG05,STE10,SCH10}, 
stabilization of cantilever oscillations in an atomic force microscope \cite{YAM09}, 
and controlling librational motion of a tethered satellite system in an elliptic orbit \cite{KOJ12}, amongst others.
For a comprehensive review of the technical implementation of the method, see \cite{PYR06,SCH08}.

In parallel to experimental realization, substantial work has been done to understand the control mechanism analytically \cite{JU97,HS05,YWH06,DHS07}.
A notable result has been reported recently in the context of refuting the alleged odd-number limitation \cite{FIE07,HA12},
believed to be a severe limitation of the delayed feedback control technique for almost a decade \cite{NAK97}. 
A way to overcome the odd-number limitation was proposed by Pyragas by using an additional unstable degree of freedom in a feedback loop
\cite{PYR01,PYR04,PYR06b}.

The original Pyragas method was subsequently improved by introducing multiple delayed signals 
into the feedback control force, both with commensurate or incommensurate 
delay times \cite{SSG94,SBG97,AP04,AP05}. Another improvement of the method was achieved by 
introducing a time-varying delay into the delayed argument of the control force which could be realized 
experimentally by changing some characteristics of the delay line in a deterministic (periodic) or a stochastic 
fashion \cite{GU08,GU10,GSU10}. For example, in coupled laser systems the delay time could be modulated by periodically changing the distance of 
the lasers or external resonator, e.g. by piezoelectric modulation. For systems acting on a slower timescale, e.g. electronic or mechanic devices, one may take advantage of digital delay lines instead, where the time-functional form of the delay modulation could be controlled by an 
external clock frequency modulator \cite{JGU12}. In this way, one could practically realize the variable-delay feedback controller and
stabilize unstable steady states and periodic orbits over a much larger domain of control parameters. 
To keep the method noninvasive also in case of unstable periodic orbits, the delay modulation should be chosen 
appropriately, e.g., in a form of a (non)periodic change of the delay between multiples of the basic period of 
the uncontrolled unstable orbit \cite{GU13}.

In this paper, we aim to provide a deeper analytical insight into the mechanism of stabilization of unstable equilibria by 
the delayed feedback method with a time-dependent delay time, and extend this concept by considering the influence of a phase-dependent coupling, the control loop latency, and multiple delay lines on the controllability, which is relevant in real experiments.
A general description of the variable-delay feedback controller is given in Sec. II, where it is shown that
the method could be explored analytically in the domain of a high-frequency modulation of the delay. 
In this case, we use the formalism of distributed delays \cite{KBS11,KBS13}, where the average contribution of the time-varying 
delay is represented by an integral kernel describing a particular delay distribution.
In Sec. III, we consider a phase-dependent coupling of the control force, which is a particular non-diagonal generalization
of the standard diagonal coupling scheme. 
As a model subject to control we use a generic two-dimensional linear system that has an unstable steady state of focus type.
This model mimics the two-dimensional center manifold of a general non-linear system, capturing the essence of the dynamics
in the vicinity of its unstable fixed point. 
In Sec. IV, we investigate the influence of control-loop latency on the efficiency of the controller. 
In Sec. V, we extend our method to the case of multiple delay feedback by considering two independent delay lines with time-varying delays.
We conclude our findings in Sec. VI.

\section{Variable-delay feedback control}

We consider a general $n$-dimensional dynamical system $\dot{\mathbf{x}}=\mathbf{F}(\mathbf{x})$, where $\mathbf{F}$ is the vector field describing the 
system's dynamics, and $\mathbf{x}=\mathbf{x}(t)\in\mathbb{R}^n$ is the state column vector. 
The equilibria $\mathbf{x}^*_i$ of the system are the solutions of $\mathbf{F}(\mathbf{x}^*_i)=0$, and the stability of a particular equilibrium $\mathbf{x}^*$
is determined by the eigenvalues of the Jacobian matrix $\widehat{\mathbf{A}}=\partial_{\mathbf{x}}{\mathbf{F}}(\mathbf{x})$ calculated at $\mathbf{x}=\mathbf{x}^*$.
In the following we assume that $\mathbf{x}^*$ is an unstable equilibrium, having at least one eigenvalue with a positive real part.

Under a linear variable-delay feedback control, the original system is transformed into
\begin{equation}
\dot{\mathbf{x}}(t)=\mathbf{F}(\mathbf{x}(t))+\widehat{\mathbf{K}}\left[\mathbf{x}(t-\tau(t))-\mathbf{x}(t)\right],
\label{contsys}
\end{equation}
where $\widehat{\mathbf{K}}$ is an $n\times n$ feedback gain matrix, and $\tau(t)$ is the time-dependent delay time. 
By choosing  $\widehat{\mathbf{K}}$ and $\tau(t)$ appropriately, we aim to stabilize the
unstable equilibrium $\mathbf{x}^*$. 
Without loss of generality, we assume that the equilibrium has already been moved to the origin by a translation of
axes, such that $\mathbf{x}^*=0$. We thus linearize the controlled system in the neighborhood of the origin to obtain 
\begin{equation}
\dot{\mathbf{x}}(t)=\widehat{\mathbf{A}} \mathbf{x}(t)+\widehat{\mathbf{K}}\left[\mathbf{x}(t-\tau(t))-\mathbf{x}(t)\right].
\label{linsys}
\end{equation}

We consider a deterministic modulation of the variable time-delay $\tau(t)$ in a fixed interval around a nominal (average) delay value $\tau_0$
\begin{equation}
\tau(t)=\tau_0+\varepsilon f(\nu t),
\label{delayvar}
\end{equation}
where $f:\mathbb{R}\rightarrow[-1,1]$ is a $2\pi$-periodic function with zero mean, and $\varepsilon$ and $\nu$ are additional variable-delay control parameters 
denoting the amplitude and the (angular) frequency of the modulation, respectively. The stability of the origin can be inferred by 
numerically integrating the linear variable-delay system (\ref{linsys}) for different values of the control parameters 
$\widehat{\mathbf{K}}$, $\tau_0$, $\varepsilon$ and $\nu$, thus determining the domains in the parameter space for which the 
stabilization can be realized.

In the limiting case of a high-frequency modulation \cite{GU10,MAN05}, when the parameter $\nu$ becomes \textit{sufficiently} large compared to the uncontrolled system's dynamics described by an intrinsic frequency $\omega$, the linearized system (\ref{linsys}) with a modulated delay has the same asymptotic stability properties as the averaged linear distributed-delay system 
\begin{equation}
\dot{\mathbf{x}}(t)=\widehat{\mathbf{A}} \mathbf{x}(t)+\widehat{\mathbf{K}}\left(\int_0^\infty \rho(\theta)\mathbf{x}(t-\theta)\;d\theta-\mathbf{x}(t)\right).
\label{ddsys}
\end{equation}
The probability density function $\rho(\theta)$, i.e. the distributed-delay kernel, is defined in a way that $\rho(\theta)d\theta$ gives the fraction of time for which $\tau(t)$ lies between $\theta$ and $\theta+d\theta$, satisfying $\rho(\theta)\geq 0$ and the probability normalization condition
\begin{equation}
\int_o^\infty\rho(\theta)\;d\theta=1.
\label{norm}
\end{equation} 
In practice, the transition to the distributed-delay limit case does not require very large modulation frequencies, and therefore 
variations of the delay in the experiment are a very practical way to create different types of delay distributions.

For a continuous modulation in an $\varepsilon$ interval around $\tau_0$ in form of a sawtooth-wave (triangular) modulation of the delay,
\begin{equation}
\tau(t)=\tau_0+\varepsilon\left[2\left(\frac{\nu t}{2\pi}\, \mathrm{mod}\, 1\right)-1\right],
\label{sawtoothdelay}
\end{equation}
$\rho(\theta)$ is a constant in the interval of the delay variation. Under the probability normalization condition (\ref{norm}) we obtain a uniform distribution
\begin{equation}
\rho(\theta)=
\begin{cases}
\frac{1}{2\varepsilon} & ,\;\theta\in[\tau_0-\varepsilon,\tau_0+\varepsilon],\\
0 & ,\;\text{elsewhere,}
\end{cases}
\label{sawmod}
\end{equation}
which does not depend on the skewness of the sawtooth function.
For a sine-wave modulation,
\begin{equation}
\tau(t)=\tau_0+\varepsilon\sin(\nu t),
\label{sinedelay}
\end{equation} 
the time interval $dt$ during which the delay $\tau$ changes by $d\tau$ is given by
\begin{equation}
dt=\frac{d\tau}{\nu\varepsilon\cos(\nu t)}=\frac{d\tau}{\nu\sqrt{\varepsilon^2-(\tau-\tau_0)^2}}.
\end{equation}
The fraction of time $dt$ within a half-period $\pi/\nu$ of the delay function $\tau$ is equivalent to the product $\rho(\tau)d\tau$. 
In terms of our previous notation, we get
\begin{equation}
\rho(\theta)=\frac{1}{\pi\sqrt{\varepsilon^2-(\theta-\tau_0)^2}}.
\label{sinmod}
\end {equation}
A periodic change of $\tau(t)$ between two fixed values $\tau_0-\varepsilon$ and $\tau_0+\varepsilon$ for the same time duration, 
\begin{equation}
\tau(t)=\tau_0+\varepsilon\,\mathrm{sgn}[\sin(\nu t)],
\label{squaredelay}
\end{equation}
results in a square wave (rectangular) modulation with a two-peak distribution
\begin{equation}
\rho(\theta)=\tfrac{1}{2} \left(\delta(\theta-\tau_0+\varepsilon)+\delta(\theta-\tau_0-\varepsilon)\right).
\label{squaremod}
\end{equation}
In an analogous way, we may obtain the corresponding distributed delay kernels for other types of delay modulations.

Applying the exponential ansatz $\mathbf{x}(t)\sim\exp(\Lambda t)$ in Eq. (\ref{ddsys}), we obtain the characteristic equation for 
the eigenvalues $\Lambda$
\begin{equation}
\det\left[\Lambda\,\widehat{\mathbf{I}}-\widehat{\mathbf{A}}+\left(1-\int_0^\infty\rho(\theta)e^{-\Lambda\theta}d\theta\right)\widehat{\mathbf{K}}\right]=0,
\label{ceq1}
\end{equation}
where $\widehat{\mathbf{I}}$ is the identity matrix. 
Since $\rho(\theta)$ is nonzero only between $\tau_0-\varepsilon$ and $\tau_0+\varepsilon$, we have
\begin{equation}
\det\left[\Lambda\,\widehat{\mathbf{I}}-\widehat{\mathbf{A}}+\left(1-e^{-\Lambda\tau_0}\chi(\Lambda,\varepsilon)\right)\widehat{\mathbf{K}}\right]=0.
\label{chareq}
\end{equation}
The quantity 
\begin{equation}
\chi(\Lambda,\varepsilon)=\int_{-\varepsilon}^{+\varepsilon}\rho(\tau_0+\theta)e^{-\Lambda\theta}d\theta
\label{chi}
\end{equation}
summarizes the effect of a given modulation, and for the above modulation types reads
\begin{equation}
\chi(\Lambda,\varepsilon)=
\begin{cases}
\frac{\sinh(\Lambda\varepsilon)}{\Lambda\varepsilon}, & \text{sawtooth-wave},\\
I_0(\Lambda\varepsilon), & \text{sine-wave},\\
\cosh(\Lambda\varepsilon), & \text{square-wave},
\end{cases}
\label{chimods}
\end{equation}
where $I_0$ is the modified Bessel function of the first kind of order 0.
In the non-modulated case, $\chi(\Lambda,0)\equiv 1$, we have the usual TDFC.
For general distributed delay, the delay term $\exp(-\Lambda \tau_0)$ is replaced in the characteristic equation (\ref{ceq1}) by the Laplace transform of the delay kernel $\rho(\theta)$ \cite{KBS11}.

\begin{figure}
\includegraphics[width=\columnwidth]{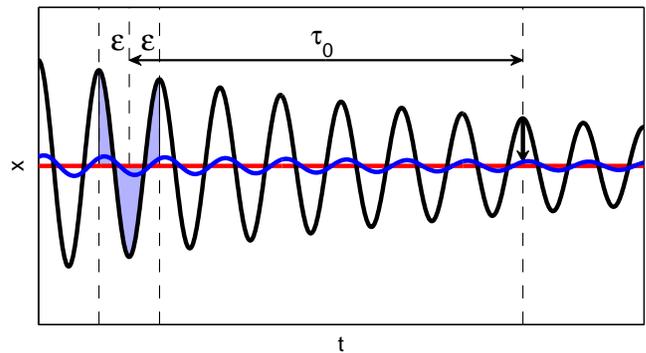}
\caption{(Color online) Sample of a trajectory $x(t)$ (black) in the vicinity of a steady state (red/gray horizontal line) of a system subjected to variable-delay feedback control with a fast sawtooth wave. The corresponding uniform delay distribution (\ref{sawmod}) creates a sliding average between $t-\tau_0-\varepsilon$ and $t-\tau_0+\varepsilon$ (area highlighted in blue/gray) which at time $t$ results in the reference signal (blue/gray, oscillating). The control force is constructed from the difference signal as marked by the vertical black arrow.}
\label{fig:interf}
\end{figure}

The function $\chi(\Lambda,\varepsilon)$ also allows for a compact explanation of the mechanism of variable-delay feedback control (VDFC). The expressions listed in Eq.~(\ref{chimods}) have regions $\{\Lambda\varepsilon\in\mathbb{C}\}$ for which $|\chi|<1$, mostly for $\mathrm{Re}(\Lambda)\ll\mathrm{Im}(\Lambda)$. In particular, values of $\chi\approx0$ can occur, meaning that the delay term effectively has been suppressed and vanishes from the characteristic Eq.~(\ref{chareq}). In the original equations of motion (\ref{ddsys}) this situation can be interpreted as \textit{destructive interference} of the delay signals covered by the distribution. Positive and negative phases of the spiral oscillation in the neighborhood of the steady state cancel each other out creating a reference term $\mathbf{x}_{\text{ref}}(t)\approx\mathbf{x}^*$ for a simplified controller
\begin{equation}
\begin{split}
\dot{\mathbf{x}}(t)&=\widehat{\mathbf{A}} \mathbf{x}(t)+\widehat{\mathbf{K}}\left(\mathbf{x}_{\text{ref}}(t)-\mathbf{x}(t)\right)\\
&\approx\widehat{\mathbf{A}} \mathbf{x}(t)+\widehat{\mathbf{K}}\left(\mathbf{x}^*-\mathbf{x}(t)\right)\;.
\end{split}
\end{equation}
The same idea can be regarded as the motivation for the original TDFC method. The advantage of VDFC lies in an improved approximation of the target state by the delay terms. Fig.~\ref{fig:interf} illustrates the control mechanism for an almost ideal situation, in which the steady state is approximated by the averaged delay signal. The instantaneous part of the coupling term stabilizes the system while pulling it towards the reference signal $\mathbf{x}_{\text{ref}}(t)$ which due to the small $\chi$-value has a smaller amplitude than the remaining instantaneous $\mathbf{x}(t)$. The resulting stabilization of the steady state shows a high robustness against parameter detuning, because the simplified control mechanism is insensitive to the phase relation between actual signal and reference signal and also works for a wide range in the coupling gain $\widehat{\mathbf{K}}$. However, the described scenario relies on a small $\chi$-value, which in general is not trivial to obtain. The delay terms can almost never be suppressed completely, so that only a rigorous consideration of the characteristic equation (\ref{chareq}) reveals the full capability of the control method. Eq.~(\ref{chareq}) is transcendental with respect to $\Lambda$, possessing an infinite set of complex solutions $\{\Lambda_i\}\in\mathbb{C}$. 
The origin can be stabilized  for those values of the control parameters $(\widehat{\mathbf{K}},\tau_0,\varepsilon)$ for which all the eigenvalues $\{\Lambda_i\}$ have negative real parts. 
The stability domain in the parameter space $(\widehat{\mathbf{K}},\tau_0,\varepsilon)$ can be calculated numerically given the modulation distribution.
Note that the control parameter $\nu$ is lost in the transition to distributed delays.

As in the usual TDFC and extended TDFC control schemes, the presence of torsion is necessary for the proposed control method to be able to stabilize equilibria. Torsion means that in its unstable subspace the steady state is only of the spiral type. To show this property, we consider the characteristic quasipolynomial $H(\Lambda)=\det\left[\Lambda\widehat{\mathbf{I}}-\widehat{\mathbf{A}}+\left(1-e^{-\Lambda\tau_0}\chi(\Lambda,\varepsilon)\right)\widehat{\mathbf{K}}\right]$.
It is easily shown that this quasipolynomial is positive for $\Lambda\rightarrow\infty$, and for $\Lambda=0$ reads
$H(0)=\det\left[-\widehat{\mathbf{A}}\right]=\prod_{n=1}^N (-e_n),$
where $e_i$ are the eigenvalues of $\widehat{\mathbf{A}}$. If $\widehat{\mathbf{A}}$ possesses an odd number of positive real eigenvalues, then $H(0)<0$, and there exists at least one positive real root of $H(\Lambda)=0$, meaning that the fixed point cannot be stabilized by the proposed method.
Although the analysis is done in the distributed-delay limit, numerical simulations show that this odd-number limitation persists also for low-frequency modulations of the delay $\tau(t)$.

The results of this section will be exploited in the following to perform a comparative analysis between the variable-delay feedback control in a distributed-delay limit and the standard delayed feedback control by using a simple normal form model as a representative of a large class of nonlinear dynamical systems. 
Specifically, we will investigate the effects of including a phase-dependent coupling matrix and the influence of nonzero latency times in the control scheme.

\section{Phase-dependent coupling}

We will apply the variable-delay feedback control to an unstable steady state of focus type. This system represents a generic model of an unstable steady state slightly above a Hopf bifurcation. In center manifold coordinates, the dynamics of the system is given by Eq. (\ref{linsys}),
where $\mathbf{x}=(x,y)^T$ is the two-dimensional column vector, and $\widehat{\mathbf{A}}$ is a $2\times2$ matrix
\begin{equation}
\widehat{\mathbf{A}}=
\left(
\begin{array}{cc}
\lambda & \omega \\
-\omega & \lambda \\
\end{array}
\right)
\label{nonperturb}
\end{equation}
that describes the dynamics in the absence of control.
Since the stability of the free-running system is determined by the eigenvalues $\lambda\pm i\,\omega$ of  
$\widehat{\mathbf{A}}$, choosing $\lambda>0$ and $\omega\neq0$ we model an unstable focus at the origin. 

In this section, we investigate a phase-dependent coupling of the control force, which is relevant, e.g., in stabilization of laser devices
where the optical phase occurs as an additional degree of freedom \cite{SCH06,DHS08}. Such couplings have also been used in the context of refuting the 
odd-number limitation of delayed feedback control in autonomous systems \cite{FIE07,SCH11,HA12}, to anticipate chaos 
synchronization \cite{PYR08}, and more recently, to control different synchronous oscillatory states of delay-coupled oscillator networks \cite{SEL12}.
The phase-dependent coupling is realized via a rotational matrix $\widehat{\mathbf{K}}$ containing a variable phase $\varphi$,
and it enters as an additional multiplicative factor to the control force
\begin{widetext}
\begin{equation}
\left(
\begin{array}{c}
\dot x(t) \\
\dot y(t) \\
\end{array}
\right) =
\left(
\begin{array}{cc}
\lambda & \omega \\
-\omega & \lambda \\
\end{array}
\right)
\left(
\begin{array}{c}
x(t) \\
y(t) \\
\end{array}
\right)+K
\left(
\begin{array}{cc}
\cos\varphi & -\sin\varphi \\
\sin\varphi & \cos\varphi \\
\end{array}
\right)
\left(
\begin{array}{c}
x(t-\tau(t))-x(t) \\
y(t-\tau(t))-y(t) \\
\end{array}
\right),
\label{phasesys}
\end{equation}
where $K$ is the feedback gain. In the distributed-delay limit, the system becomes
\begin{equation}
\left(
\begin{array}{c}
\dot x(t) \\
\dot y(t) \\
\end{array}
\right) =
\left(
\begin{array}{cc}
\lambda & \omega \\
-\omega & \lambda \\
\end{array}
\right)
\left(
\begin{array}{c}
x(t) \\
y(t) \\
\end{array}
\right)+K
\left(
\begin{array}{cc}
\cos\varphi & -\sin\varphi \\
\sin\varphi & \cos\varphi \\
\end{array}
\right)
\left(
\begin{array}{c}
\int_0^\infty \rho(\theta)x(t-\theta)\;d\theta-x(t) \\
\int_0^\infty \rho(\theta)y(t-\theta)\;d\theta-y(t) \\
\end{array}
\right),
\label{phasedd}
\end{equation}
and the stability of the origin is determined by the roots $\Lambda$ of the characteristic equation
\begin{equation}
\left[\Lambda-\lambda+K\left(1-e^{-\Lambda\tau_0}\chi(\Lambda,\varepsilon)\right)\cos\varphi\right]^2+\left[\omega+K\left(1-e^{-\Lambda\tau_0}\chi(\Lambda,\varepsilon)\right)\sin\varphi\right]^2=0.
\label{charphio}
\end{equation}
\end{widetext}
This equation can be further simplified to
\begin{equation}
\Lambda+Ke^{\mp i\varphi}\left(1-e^{-\Lambda\tau_0}\chi(\Lambda,\varepsilon)\right)=\lambda\pm i\omega.
\label{charphi}
\end{equation}
The control method is successful if there exists a nonempty set in the four-dimensional parameter space $(K,\tau_0,\varepsilon,\varphi)$ for which the real parts of all the characteristic roots $\Lambda$ are negative. 

\begin{figure}
\includegraphics[width=\columnwidth,height=!]{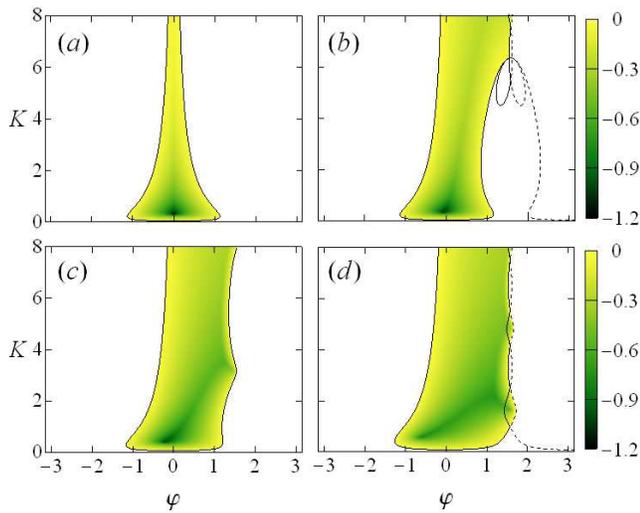}
\caption{(Color online) Control domain in the $(\varphi,K)$ plane for a sawtooth-wave
modulation of the delay for different modulation amplitudes: (a) $\varepsilon=0$, 
(b) $\varepsilon=0.3$, (c) $\varepsilon=0.5$ and (d) $\varepsilon=0.9$. 
The nominal delay is fixed at $\tau_0=1$, which is optimal value in the non-modulated
case (TDFC). The parameters of the free-running system are $\lambda=0.1$ and
$\omega=\pi$. The grayscale (color code) depicts only those values of the control parameters
for which the largest real part of the complex eigenvalues $\Lambda$ is negative, indicating
a successful control. The depicted solid lines that enter into a description of the stability boundary are
calculated from Eqs. (\ref{kphi}) and (\ref{phiphi1}), and the dashed lines from Eqs. (\ref{kphi}) and (\ref{phiphi2}). 
For clarity of the picture, we depict only a few branches of the boundary curves.}
\label{fig:sawphiK}
\end{figure}

\begin{figure}
\includegraphics[width=\columnwidth,height=!]{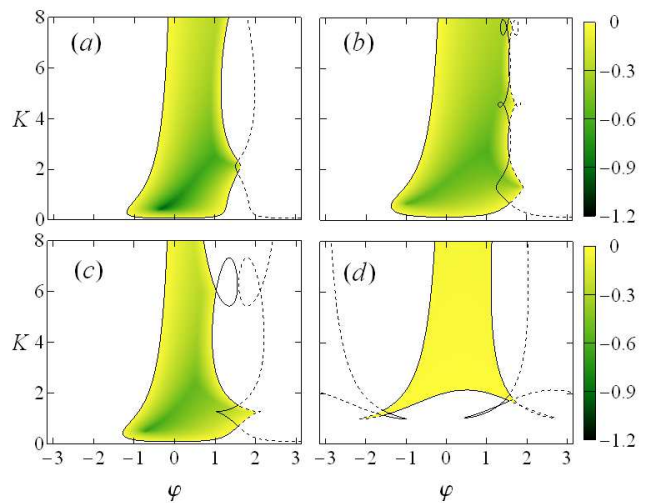}
\caption{(Color online) 
Same as Fig.~\ref{fig:sawphiK} for a sine-wave modulation of the delay: (a) $\varepsilon=0.5$, (b) $\varepsilon=0.9$, and for a square-wave modulation: (c) $\varepsilon=0.5$, (d) $\varepsilon=0.9$. Other parameters are the same as in Fig.~\ref{fig:sawphiK}.}
\label{fig3}
\end{figure}

The domain of control in the plane parametrized by the feedback phase $\varphi$ and the feedback gain $K$ is given in Figs. \ref{fig:sawphiK} and \ref{fig3}.
The grayscale (color code) indicates only those control parameters $(\varphi,K)$ for which the leading eigenvalues have negative real parts,
and the control is more robust as these values are more negative. 
Panels (a)-(d) in Fig. \ref{fig:sawphiK} depict control domains for a sawtooth-wave modulation corresponding to a fixed nominal delay $\tau_0=1$ and different modulation amplitudes $\varepsilon$: (a) $\varepsilon=0$, (b) $\varepsilon=0.3$, (c) $\varepsilon=0.5$ and (d) $\varepsilon=0.9$. The parameters of the unstable focus are fixed at $\lambda=0.1$ and $\omega=\pi$.
The chosen nominal delay $\tau_0=1$ is an optimal value in the non-modulated control case and $\varphi=0$ (diagonal coupling).
From the corresponding control domains, it is observed that an increase of the modulation amplitude $\varepsilon$
leads to an enlargement of the control domain in the direction of the positive $\varphi$-axis, thus destroying the symmetry of the stability island centered
at $\varphi=0$. A similar result is observed for a sine-wave modulation in panels (a)-(b) in Fig. \ref{fig3}, corresponding
to $\varepsilon=0.5$ (a) and $\varepsilon=0.9$ (b).
A different behavior is observed in panels (c)-(d) in Fig. \ref{fig3} depicting the corresponding stability domains for a square-wave modulation. 
In the latter case, the stability domain increases more rapidly as $\varepsilon$ is increased from zero, until $\varepsilon$ reaches a critical value after 
which the domain decreases, resulting in an instability stripe at $\varphi=0$ when $\varepsilon=1$. 
This non-monotonic change of the stability domain is typical for a square-wave modulation, and it has
been observed in other circumstances. 

In order to understand the mechanisms behind these numerical results, we first recall the intuitive argument about destructive interference and the simplified controller from the previous section. From Eq.~(\ref{phasedd}) we directly see that the stabilizing diagonal elements of the instantaneous control term are weighted with $\cos\varphi$. Since the onset of stability $K_{\mathrm{min}}$ is dominated by the competition between $\lambda>0$ and $-K\cos\varphi<0$, we expect $K_{\mathrm{min}}\propto\lambda/\cos\varphi$, see also the detailed reasoning in Eq.~(\ref{kminphi2}) below. For $|\varphi|>\pi/2$ the instantaneous part of the coupling becomes repelling. Therefore with regard to the depicted control mechanism we expect the control to largely fail in this region. All simulations clearly support this picture. Small exceptional regions of stability exceeding this boundary arise from the non-vanishing delay terms which can support stabilization given an optimal parameter constellation. However, for large values of the mean delay $\tau_0$ this type of support fails and $|\varphi|=\pi/2$ becomes a strict limit of control.

\begin{figure}
\includegraphics[width=\columnwidth]{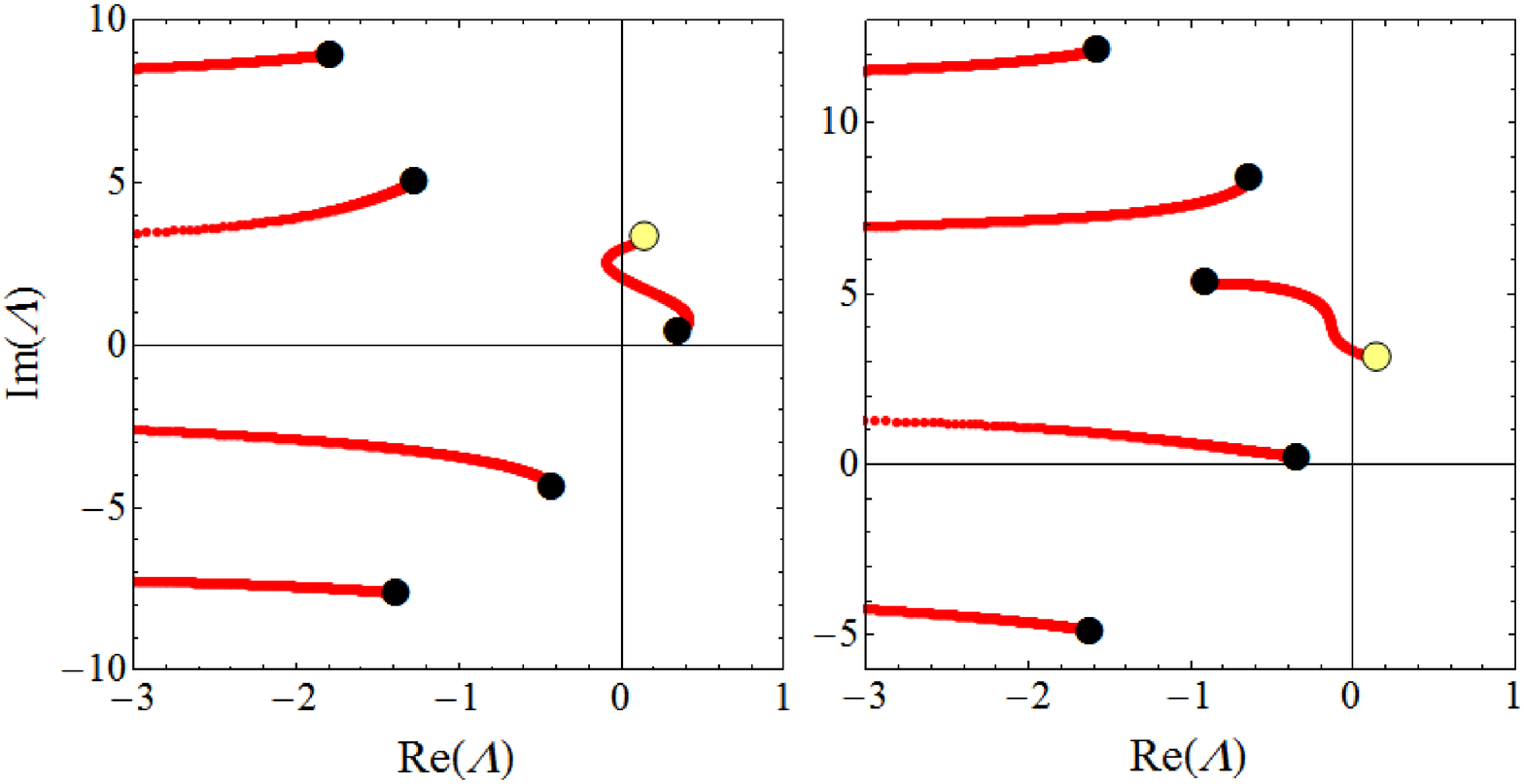}
\caption{(Color online) Characteristic exponents $\Lambda$ in the complex plane parametrized by the feedback gain $K$. Left: $\varphi=-1$, right: $\varphi=+1$. Depicted are five leading eigenvalue curves for $K\in[0,6]$. The gray-filled circle (yellow online) corresponds to $K=0$ and the black dots correspond to $K=6$.  The eigenmodes created by the variable-delay control originate from $\mathrm{Re}(\Lambda)=-\infty$. Other parameters as in Fig.~\ref{fig:sawphiK}c. }
\label{fig:LambdainC}
\end{figure}

The asymmetry can be explained by the interaction of the non-diagonal elements of system terms and control terms in Eq.~(\ref{phasedd}). If $\omega$ were zero, the control term would have no preference for positive or negative values of $\varphi$ because the change of $\varphi\rightarrow-\varphi$ would be equivalent to mirroring the complete system at one coordinate axis, e.g. $x\rightarrow-x$. A non-zero value of the spiral frequency $\omega$ breaks the symmetry and leads to different resulting frequencies $\Omega=\mathrm{Im}(\Lambda)$ under control for different signs of $\varphi$. From the interference mechanism one can conclude, that at least for the continuous distributions (\ref{sawmod}) and (\ref{sinmod}) high frequencies are more favorable for successful control, because a low $\chi$-value corresponds to many oscillation periods covered by the distribution. Indeed, if we have a closer look at the parameter constellation underlying Fig.~\ref{fig:sawphiK}, we see that for negative $\varphi$ the imaginary part of the most unstable mode tends to vanish, thus inhibiting the interference effect and leading to control failure. In contrast, for positive $\varphi$ the frequency increases with increasing $K$, thus improving control efficiency. A parametric plot of different $\Lambda_i$ in the complex plane reveals this property, see Fig.~\ref{fig:LambdainC}.

A detailed analytical investigation of the control domain can be done by using the characteristic equation (\ref{charphi}) and noticing that
the transition from instability to stability occurs at the boundaries of the control domain, where the leading eigenvalues $\Lambda$ are purely imaginary. 
Setting $\Lambda=i\Omega$ in Eq. (\ref{charphi}), and separating real and imaginary parts results in a system of two real-valued equations
\begin{eqnarray}
K\cos\varphi-K\chi(i\Omega,\varepsilon)\cos(\Omega\tau_0\pm\varphi)&=&\lambda,
\label{sys1}\\
\Omega\mp K\sin\varphi+K\sin(\Omega\tau_0\pm\varphi)\chi(i\Omega,\varepsilon)&=&\pm\omega,
\label{sys2}
\end{eqnarray}
which is an implicit parametric representation of the boundaries of the control domain in the parameter 
space $(K,\tau_0,\varepsilon,\varphi)$, in which the eigenfrequency $\Omega$ has the role
of a parametrization variable. 
We have taken into account that for the above mentioned distribution kernels, $\rho(\theta)$ is even around the nominal delay $\tau_0$, meaning that $\chi(i\Omega,\varepsilon)$ is a real-valued function
\begin{equation}
\chi(i\Omega,\varepsilon)=
\begin{cases}
\frac{\sin(\Omega\varepsilon)}{\Omega\varepsilon}, & \text{sawtooth-wave},\\
J_0(\Omega\varepsilon), & \text{sine-wave},\\
\cos{(\Omega\varepsilon)}, & \text{square-wave},
\end{cases}
\label{chireal}
\end{equation}
where $J_0$ is the Bessel function of the first kind of order 0.
Considering the positivity of the nominal delay $\tau_0$ and the feedback gain $K$, as well as the positivity of the parameters $\varphi,\lambda$ and $\omega$
used in the current analysis, elimination of the phase parameter $\varphi$ from the last system of transcendental
equations leads to an expression for $K$ in terms of the eigenfrequency $\Omega$
\begin{equation}
K(\Omega)=\sqrt{\frac{\lambda^2+(\omega-\Omega)^2}{\sin^2(\Omega\tau_0)+[\chi(i\Omega,\varepsilon)-\cos(\Omega\tau_0)]^2}}.
\label{kphi}
\end{equation} 
Taking into account the multivalued properties of the arcsine function, one obtains in a similar manner the analytical expressions for the phase parameter $\varphi$ in dependence on $\Omega$
\begin{widetext}
\begin{eqnarray}
\varphi_1(\Omega)&=&\arcsin\left({K(\Omega)\frac{\chi(i\Omega,\varepsilon)[\lambda\sin(\Omega\tau_0)+(\omega-\Omega)\cos(\Omega\tau_0)]-(\omega-\Omega)}{\lambda^2+(\omega-\Omega)^2}}\right)+2n\pi \label{phiphi1}\\
\varphi_2(\Omega)&=&-\arcsin\left({K(\Omega)\frac{\chi(i\Omega,\varepsilon)[\lambda\sin(\Omega\tau_0)+(\omega-\Omega)\cos(\Omega\tau_0)]-(\omega-\Omega)}{\lambda^2+(\omega-\Omega)^2}}\right)+(2n+1)\pi,
\label{phiphi2}
\end{eqnarray} 
\end{widetext}
where $n$ is a non-negative integer. Equations (\ref{kphi})--(\ref{phiphi2}) describe the boundary of the control domain in Figs. \ref{fig:sawphiK} and \ref{fig3}, where the solid and dashed lines represent branches of $\varphi_1(\Omega)$ and $\varphi_2(\Omega)$, respectively.

In the case of zero feedback phase ($\varphi=0$), the phase-dependent feedback force is reduced to a diagonal coupling. In the absence of modulation (TDFC), the domain of control in the plane 
spanned by the feedback gain and the nominal delay consists of several distinct stability islands centered at odd values of $\tau_0$, 
separated by regions corresponding to even $\tau_0$ for which the stabilization fails for any $K$. This observation can be inferred from Eqs. (\ref{sys1})-(\ref{sys2})
setting $\varphi=0$ and $\varepsilon=0$. When $\Omega \tau_0=(2n+1)\pi$, then $K_\mathrm{min}=\lambda/2$ and $\Omega=\omega$. On the other hand, when $\Omega \tau_0=2n\pi$, 
the control fails for any finite value of $K$. A detailed analysis of the control boundaries in this case (TDFC) is provided in Ref. \cite{HS05}. 
By the same reasoning, in the presence of modulation ($\varepsilon>0$) and zero phase, the corresponding minimum feedback gain values are 
\begin{equation}
K_\mathrm{min}(\varepsilon)=
\begin{cases}
\frac{\displaystyle \lambda}{\displaystyle [1+\chi(i\omega,\varepsilon)]}, & \tau_0=(2n+1)\pi/\omega,\\\\
\frac{\displaystyle \lambda}{\displaystyle [1-\chi(i\omega,\varepsilon)]}, & \tau_0=2n\pi/\omega,
\end{cases}
\label{kminvdfc}
\end{equation} 
leading to a reconfiguration of the control domain depending on the type of the delay modulation \cite{GU08}. Specifically, it is observed that the instability stripes at even values of $\tau_0$ cease to exist, and the stability islands are starting to interconnect as soon as $\varepsilon>0$. Although for a sawtooth and sine-wave modulation the enlargement of the control domain with increasing $\varepsilon$ is monotonic, an oscillatory behavior is observed for a square-wave modulation due to the form of the $\chi$ function (i.e. $\cos(\omega\varepsilon)$) in the denominator of the expressions for $K_\mathrm{min}$. In the latter case, the stability islands centered at $\tau_0=(2n+1)\pi/\omega$ are first enlarged and eventually interconnected with increasing $\varepsilon$, up to some $\varepsilon$ value after which the control domain shrinks, gradually collapsing into several stability islands centered at $\tau_0$ equal to even multiples of $\pi/\omega$. 

For $\varphi\neq0$ and a variable-delay control force ($\varepsilon>0$), the minimum feedback gain at the above specific values of $\tau_0$  cannot be deduced from Eqs. (\ref{sys1})-(\ref{sys2}) by following the previous lines of deduction. Nevertheless, we can make approximate predictions for small values of $\varphi$, when $\Omega\tau_0$ is an integer multiple of $\pi$, and we have $\Omega\approx\omega$ from Eq. (\ref{sys2}). Hence, in the regime of small values of $\varphi$, the minimum feedback gain is
\begin{equation}
K_\mathrm{min}(\varepsilon,\varphi)=
\begin{cases}
\frac{\displaystyle \lambda}{\displaystyle [1+\chi(i\omega,\varepsilon)]\cos\varphi}, & \tau_0\approx (2n+1)\pi/\omega,\\\\
\frac{\displaystyle \lambda}{\displaystyle [1-\chi(i\omega,\varepsilon)]\cos\varphi}, & \tau_0\approx 2n\pi/\omega.
\end{cases}
\label{kminphi2}
\end{equation}  
The resulting expressions for $K_\mathrm{min}$ show that in the absence of any delay modulation and for small $\varphi$, the minimum feedback gain at the "optimal" delay values $\tau_0=(2n+1)\pi/\omega$ is $K_\mathrm{min}=\lambda/(2\cos\varphi)$, containing the dependence on $\varphi$ via $\cos\varphi$ in the denominator. Consequently, the principal stability island in $(\varphi,K)$ parametric plane is centered at $\varphi=0$, for which $K_\mathrm{min}=\lambda$ (see panel (a) in Fig. \ref{fig:sawphiK}). On the other hand, at $\tau_0=2n\pi/\omega$, the stabilization cannot be achieved for any $K$, regardless of the value of the phase parameter $\varphi$. 
For a non-zero modulation amplitude $\varepsilon$ at small $\varphi$, the values of $K_\mathrm{min}$ depend on the type and the amplitude of the delay modulation.  
Although the form of $\chi(i\omega,\varepsilon)$ for sawtooth and sine-wave modulations is such that the control in the variable-delay case is generally possible for all nominal delays $\tau_0\geq\varepsilon$, this is not the case for a square-wave modulation. Namely, in the latter case, $\chi(i\omega,\varepsilon)=\cos(\omega\varepsilon)$, giving a non-monotonic behavior of $K_\mathrm{min}$ with increasing modulation amplitude $\varepsilon$. Specifically, for $\varepsilon=(2n+1)\pi/\omega$, the control fails at $\tau_0=(2n+1)\pi/\omega$ for any $K$, but it is optimal at $\tau_0=2n\pi/\omega$.   

\begin{figure}
\includegraphics[width=\columnwidth,height=!]{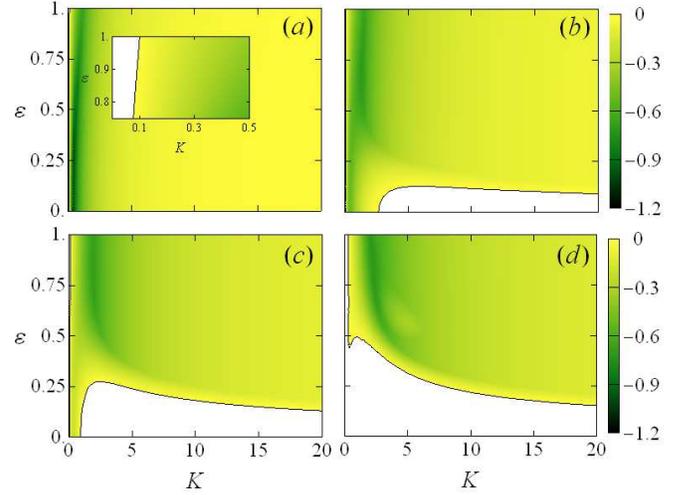}
\caption{(Color online) Control domain in the $(K,\varepsilon)$ plane for a sawtooth-wave
modulation of the delay for different values of the feedback phase $\varphi$: (a) $\varphi=0$, 
(b) $\varphi=\pi/8$, (c) $\varphi=\pi/4$ and (d) $\varphi=3\pi/8$. 
The nominal delay is fixed at $\tau_0=1$. The parameters of the free-running system are as in Fig. \ref{fig:sawphiK}.}
\label{fig5}
\end{figure}

\begin{figure}
\includegraphics[width=\columnwidth,height=!]{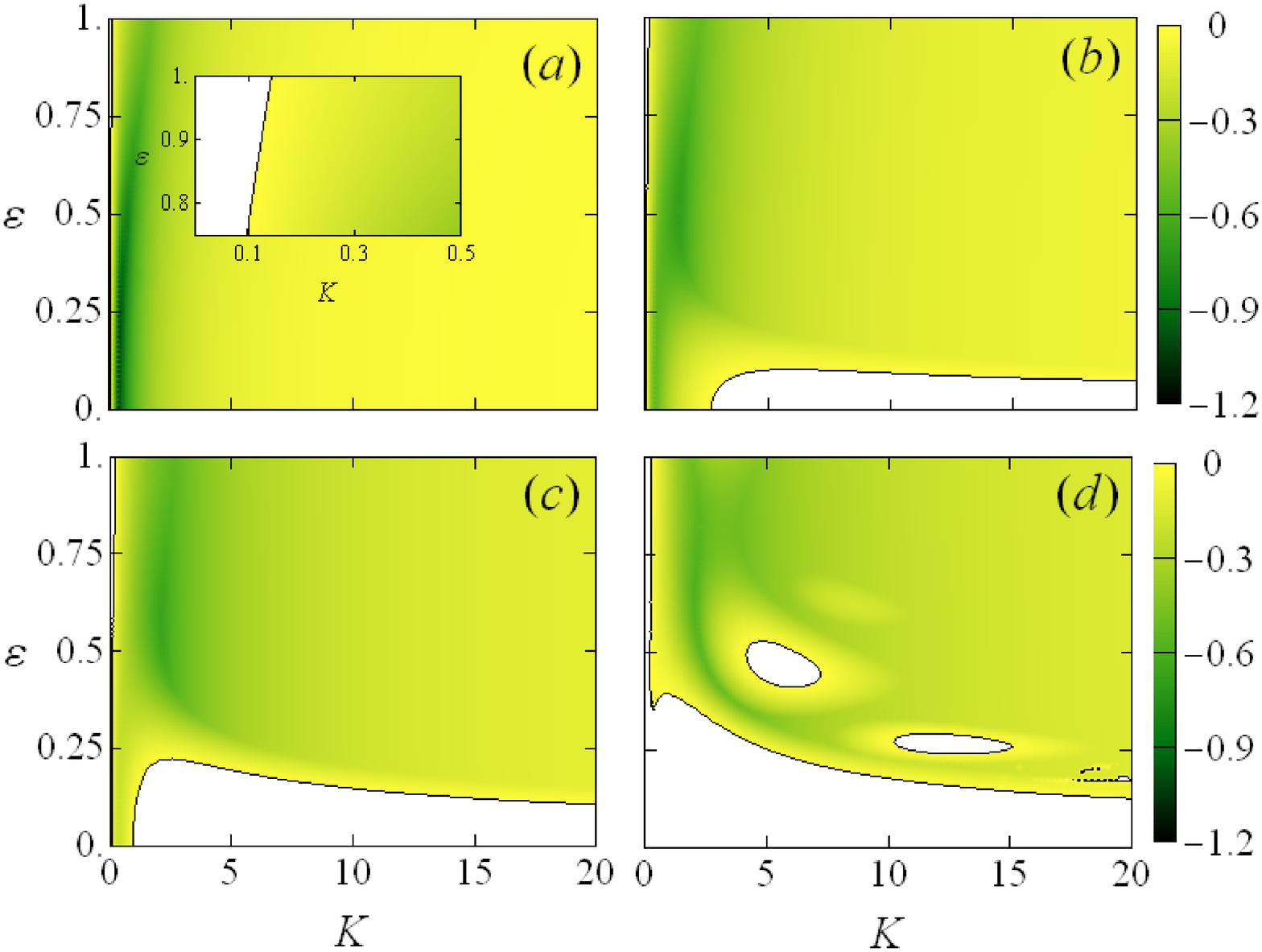}
\caption{(Color online) The corresponding control domain in the $(K,\varepsilon)$ plane for a sine-wave
modulation of the delay for different values of the feedback phase $\varphi$: (a) $\varphi=0$, 
(b) $\varphi=\pi/8$, (c) $\varphi=\pi/4$ and (d) $\varphi=3\pi/8$. 
The other parameters are as in Fig. \ref{fig5}.}
\label{fig6}
\end{figure}

\begin{figure}
\includegraphics[width=\columnwidth,height=!]{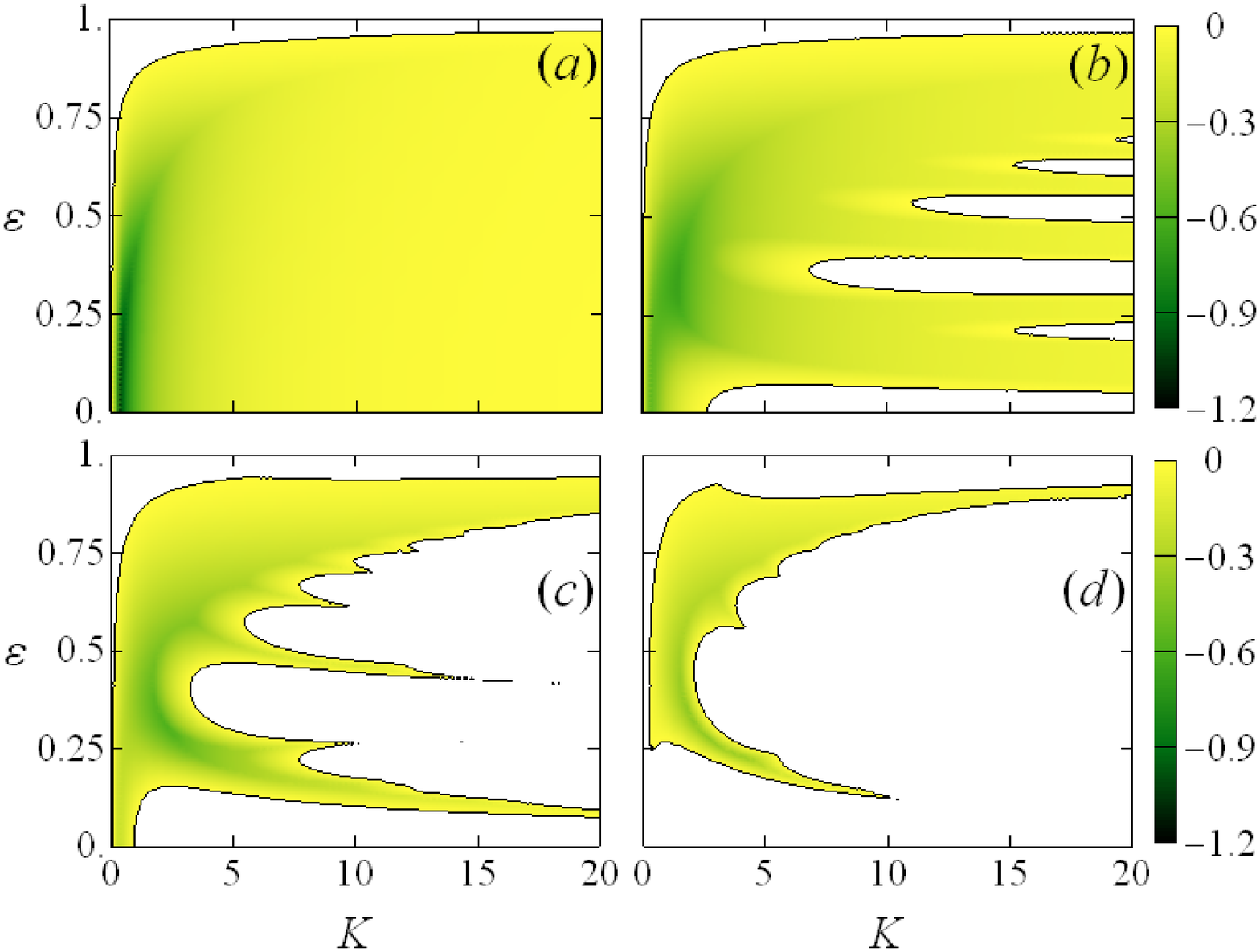}
\caption{(Color online) The control domain in the $(K,\varepsilon)$ plane for a square-wave
modulation of the delay for different values of the feedback phase $\varphi$: (a) $\varphi=0$, 
(b) $\varphi=\pi/8$, (c) $\varphi=\pi/4$ and (d) $\varphi=3\pi/8$. 
The other parameters are as in Fig. \ref{fig5}.}
\label{fig7}
\end{figure}

To further investigate the effects of the modulation amplitude $\varepsilon$ on the control efficiency, we have calculated the stability domains in the parametric plane $(K,\varepsilon)$ for different modulation types and different values of the phase parameter $\varphi$. The results are depicted in Figs. \ref{fig5}, \ref{fig6} and \ref{fig7} for sawtooth-, sine- and square-wave modulation, respectively. Different panels in each figure correspond to different phases: (a) $\varphi=0$, (b) $\varphi=\pi/8$, (c) $\varphi=\pi/4$, (d) $\varphi=3\pi/8$. 
The parameters of the unperturbed system are $\lambda=0.1$ and $\omega=\pi$, and the nominal delay is fixed at $\tau_0=1$ as before. As the modulation amplitude increases from zero, a considerable reconfiguration of the stability domain is observed, depending on the type of the delay modulation. Increasing the modulation amplitude enlarges the interval of $K$ for successful control, although the enlargement is not necessarily monotonic, and the domain may consist of several disconnected intervals. The improvement of the delayed feedback controller from including time-varying delay is evident from the diagrams. For example, for a phase parameter value $\varphi=3\pi/8$ (panel (d) in Figs. \ref{fig5}, \ref{fig6} and \ref{fig7}), while the original Pyragas method fails for any $K$, the variable-delay feedback method is able to stabilize the fixed point for certain values of the modulation amplitude $\varepsilon>0$. 

The parametric representation of the stability boundaries in $(K,\varepsilon)$ plane can be obtained from Eqs. 
(\ref{sys1}) and (\ref{sys2})
\begin{eqnarray}
K(\Omega)&=&\frac{\lambda\sin(\Omega\tau_0+\varphi)+(\omega-\Omega)\cos(\Omega\tau_0+\varphi)}{\sin(\Omega\tau_0)},\nonumber\\ \\
\chi(i\Omega,\varepsilon)&=&\frac{\lambda\sin\varphi+(\omega-\Omega)\cos\varphi}{\lambda\sin(\Omega\tau_0+\varphi)+(\omega-\Omega)\cos(\Omega\tau_0+\varphi)}.\nonumber\\
\end{eqnarray}
In the second equation, $\varepsilon$ is implicitly contained in $\chi(i\Omega,\varepsilon)$, which is a real function in the delay modulation cases
considered above (see Eq. \ref{chireal}).
In general it is not possible to invert the function $\chi$ analytically. One should treat the last system as implicit parametric representation of the control boundaries and seek for the stability curves numerically. The boundaries of the corresponding control domains are given in Figs. \ref{fig5}--\ref{fig7} by the solid line.

\section{Control-loop latency}

In an experimental realization of the control method, one has to take into account the latency of the feedback 
circuit due to the time required for the generation of the feedback control signal and its reinjection into the system. 
In laser systems, the latency is associated with the time the light takes to traverse the distance between the laser and the Fabry-Perot controller.
It has been shown that latency time always acts destructively upon the control domains, reducing the effectiveness of the
controller \cite{JUS99,HS03,HS05,YWH06,DHS08}.

\begin{figure}
\includegraphics[width=\columnwidth,height=!]{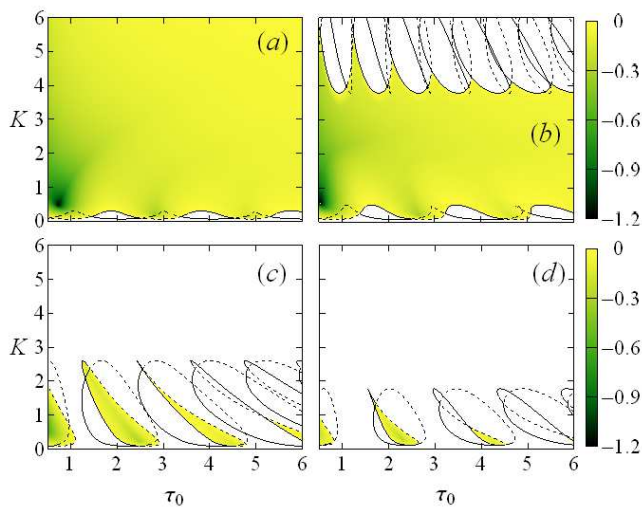}
\caption{(Color online) Control domain in the $(\tau_0,K)$ plane for a sawtooth-wave
modulation of the delay for different values of the 
latency time: (a) $\delta=0.1$, (b) $\delta=0.2$, (c) $\delta=0.3$ and (d) $\delta=0.4$. 
The modulation amplitude is fixed at $\varepsilon=0.5$, for which the stability domain fills almost the entire depicted
part of the parametric plane in the case of zero latency.  The depicted solid lines that enter into a description of the stability boundary are
calculated from Eqs. (\ref{klat}) and (\ref{taulat1}), and the dashed lines from Eqs. (\ref{klat}) and (\ref{taulat2}).
The parameters of the free-running system are as in Fig. \ref{fig:sawphiK} and $\varphi=0$.}
\label{fig8}
\end{figure}

\begin{figure}
\includegraphics[width=\columnwidth,height=!]{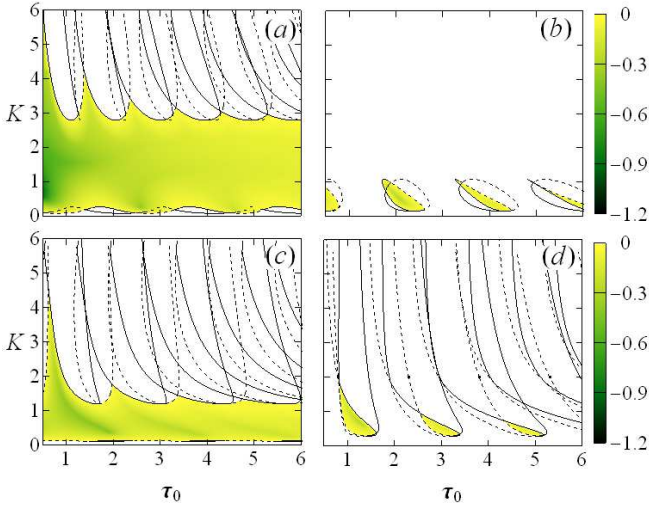}
\caption{(Color online) 
Same as Fig. \ref{fig8} for a sine-wave modulation of the delay for fixed  $\varepsilon=0.5$ and different values of the latency time: (a) $\delta=0.2$, (b) $\delta=0.4$, and for a square-wave modulation: (c) $\delta=0.2$, (d) $\delta=0.4$. Other parameters are the same as in Fig. \ref{fig8}.}
\label{fig9}
\end{figure}

In this section, we investigate the effects of latency on the variable-delay feedback control in the distributed-delay limit.
The latency time $\delta$ enters as an additional constant time-delay in the control force, and the system now reads
\begin{widetext}
\begin{equation}
\left(
\begin{array}{c}
\dot x(t) \\
\dot y(t) \\
\end{array}
\right) =
\left(
\begin{array}{cc}
\lambda & \omega \\
-\omega & \lambda \\
\end{array}
\right)
\left(
\begin{array}{c}
x(t) \\
y(t) \\
\end{array}
\right)+K
\left(
\begin{array}{c}
x(t-\delta-\tau(t))-x(t-\delta) \\
y(t-\delta-\tau(t))-y(t-\delta) \\
\end{array}
\right).
\label{latsys}
\end{equation}
In the distributed-delay limit, the system becomes
\begin{equation}
\left(
\begin{array}{c}
\dot x(t) \\
\dot y(t) \\
\end{array}
\right) =
\left(
\begin{array}{cc}
\lambda & \omega \\
-\omega & \lambda \\
\end{array}
\right)
\left(
\begin{array}{c}
x(t) \\
y(t) \\
\end{array}
\right)+K
\left(
\begin{array}{c}
\int_0^\infty \rho(\theta)x(t-\delta-\theta)\;d\theta-x(t-\delta) \\
\int_0^\infty \rho(\theta)y(t-\delta-\theta)\;d\theta-y(t-\delta) \\
\end{array}
\right),
\label{latdd}
\end{equation}
\end{widetext}
leading to the characteristic equation
\begin{equation}
\left[\Lambda-\lambda+Ke^{-\Lambda\delta}\left(1-e^{-\Lambda\tau_0}\chi(\Lambda,\varepsilon)\right)\right]^2+\omega^2=0,
\label{latlat}
\end{equation}
that can be further simplified into
\begin{equation}
\Lambda+Ke^{-\Lambda\delta}\left(1-e^{-\Lambda\tau_0}\chi(\Lambda,\varepsilon)\right)=\lambda\pm i\omega.
\label{charlat}
\end{equation}
Upon separating the real and imaginary parts, one obtains a system of two real-valued equations. 
We again consider the control boundary given by $\Lambda=i \Omega$ which is given by the implicit parametric representation
\begin{eqnarray}
K\cos(\Omega\delta)-K\cos(\Omega(\delta+\tau_0))\chi(i\Omega,\varepsilon)&=&\lambda,
\label{latsys1}\\
\Omega-K\sin(\Omega\delta)+K\sin(\Omega(\delta+\tau_0))\chi(i\Omega,\varepsilon)&=&\pm\omega.
\label{latsys2}
\end{eqnarray}
Figure \ref{fig8} depicts the domain of control in the $(\tau_0,K)$ parametric plane for a sawtooth-wave modulation at a fixed modulation amplitude $\varepsilon=0.5$ and increasing latency time $\delta$: (a) $\delta=0.1$, (b) $\delta=0.2$, (c) $\delta=0.3$, (d) $\delta=0.4$. The control domains for a sine-wave modulation corresponding to $\delta=0.2$ and $\delta=0.4$ are depicted in panels (a) and (b) in Fig. \ref{fig9}, and panels (c) and (d) are related domains for a square-wave modulation. The parameter values of the uncontrolled system are set to $\lambda=0.1$ and $\omega=\pi$ as before. For a zero latency time, the control domain fills the whole depicted range of the parametric plane down to some minimum value of $K$, which for $\tau_0$ equal to an integer multiple of $\pi/\omega$ is explicitly given by Eq. (\ref{kminvdfc}).
It is observed that increasing latency time reduces the area and the robustness of the control domain.  
In analogy to Sec. II, one can derive approximate expressions for $K_\mathrm{min}$ for small values of the latency $\delta$
\begin{equation}
K_\mathrm{min}(\varepsilon,\delta)=
\begin{cases}
\frac{\displaystyle \lambda}{\displaystyle [1+\chi(i\omega,\varepsilon)]\cos(\omega\delta)}, & \tau_0=(2n+1)\pi/\omega,\\\\
\frac{\displaystyle \lambda}{\displaystyle [1-\chi(i\omega,\varepsilon)]\cos(\omega\delta)}, & \tau_0=2n\pi/\omega.
\end{cases}
\label{kminphi}
\end{equation}  
In parallel to the conclusions of the previous section, inclusion of a variable time-delay in the feedback control force generally enlarges the control domain, making the control possible even for those values of $\tau_0$ for which the control always fails at any $K$ in the non-modulated case.

To describe the boundaries of the control domain in the ($\tau_0$,K) plane analytically, we algebraically manipulate the system (\ref{latsys1})-(\ref{latsys2}) 
and find two families of branches of solutions for $K$ and $\tau_0$ parametrized by the eigenfrequency $\Omega$
\begin{widetext}
\begin{eqnarray}
K(\Omega)&=&\frac{\lambda\cos(\Omega\delta)-(\omega-\Omega)\sin(\Omega\delta)\pm\sqrt{[\lambda\cos(\Omega\delta)-(\omega-\Omega)\sin(\Omega\delta)]^2-(1-\chi^2(i\Omega,\varepsilon))[\lambda^2+(\omega-\Omega)^2]}}{1-\chi^2(i\Omega,\varepsilon)},\label{klat}\\
\tau_{0,1}(\Omega)&=&\frac{1}{\Omega}\left[\arcsin\left(\frac{\lambda\sin(\Omega\delta)+(\omega-\Omega)\cos(\Omega\delta)}{K(\Omega)\chi(i\Omega,\varepsilon)}\right)+2n\pi\right],\label{taulat1}\\
\tau_{0,2}(\Omega)&=&\frac{1}{\Omega}\left[-\arcsin\left(\frac{\lambda\sin(\Omega\delta)+(\omega-\Omega)\cos(\Omega\delta)}{K(\Omega)\chi(i\Omega,\varepsilon)}\right)+(2n+1)\pi\right],
\label{taulat2}
\end{eqnarray}
\end{widetext}
where $n$ is a non-negative integer that takes care of the different branches due to the multivalued arcsine function involved in the boundary description.
The control domain boundaries in Figs. \ref{fig8} and \ref{fig9} are represented parametrically by Eqs. (\ref{klat})--(\ref{taulat2}), where the branches related to $\tau_{0,1}$ are given by the solid lines, and the branches of $\tau_{0,2}$ with dashed lines.

\begin{figure}
\includegraphics[width=\columnwidth,height=!]{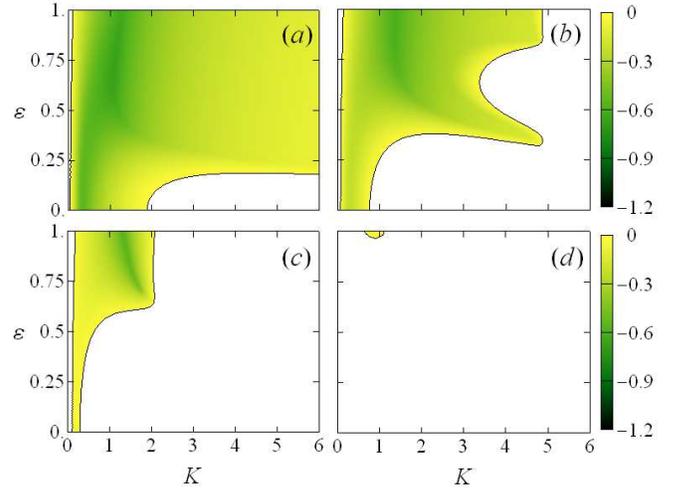}
\caption{(Color online) Control domain in the $(K,\varepsilon)$ plane for a sawtooth-wave
modulation of the delay for different values of the latency time: (a) $\delta=0.1$, (b) $\delta=0.2$, (c) $\delta=0.3$ and (d) $\delta=0.4$. 
The nominal time delay is fixed at $\tau_0=1$. The parameters of the free-running system are as in Fig. \ref{fig:sawphiK}.}
\label{fig10}
\end{figure}

\begin{figure}
\includegraphics[width=\columnwidth,height=!]{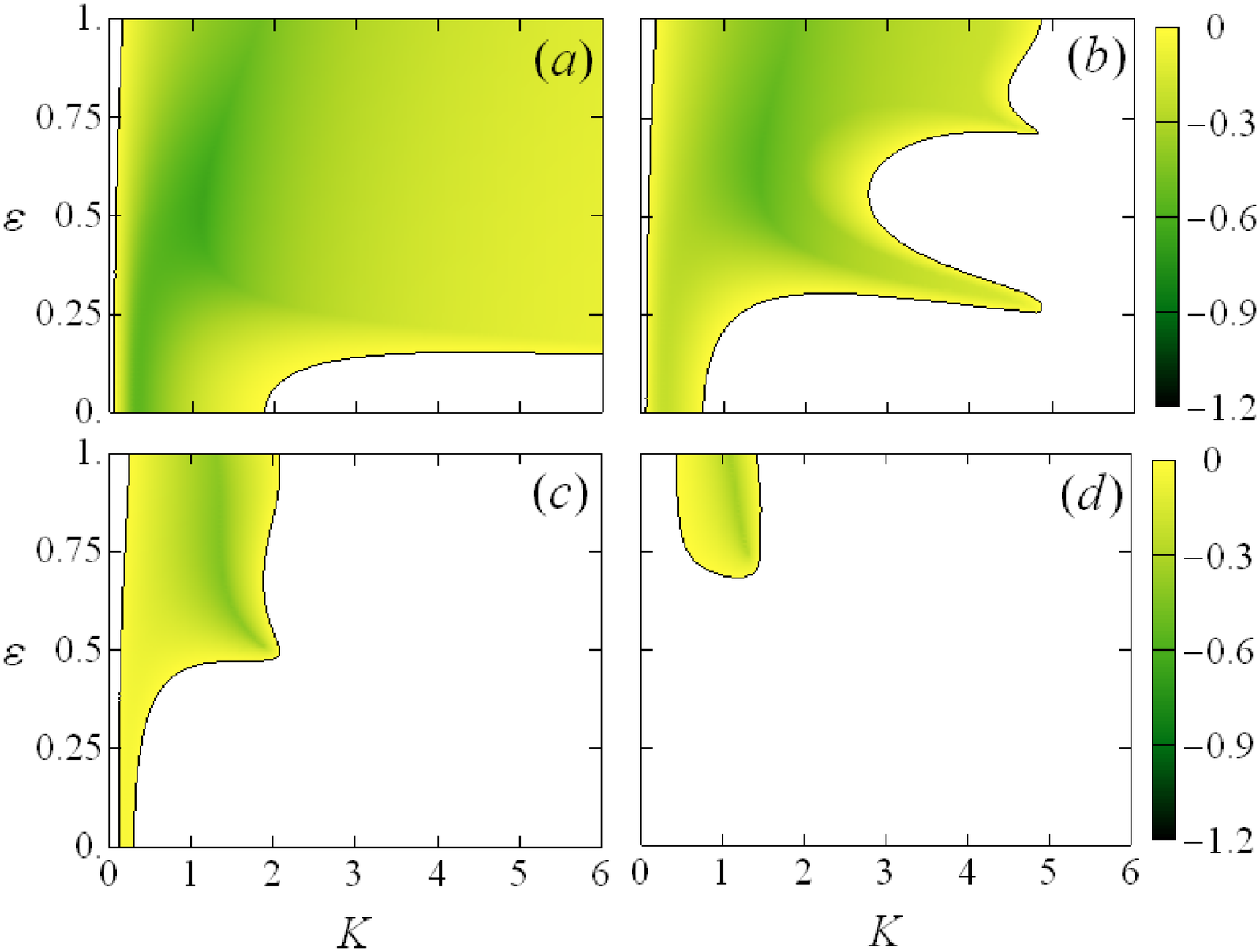}
\caption{(Color online) Control domain in the $(K,\varepsilon)$ plane for a sine-wave
modulation of the delay for different values of the latency time: (a) $\delta=0.1$, (b) $\delta=0.2$, (c) $\delta=0.3$ and (d) $\delta=0.4$. 
The other parameters are as in Fig. \ref{fig10}.}
\label{fig11}
\end{figure}

\begin{figure}
\includegraphics[width=\columnwidth,height=!]{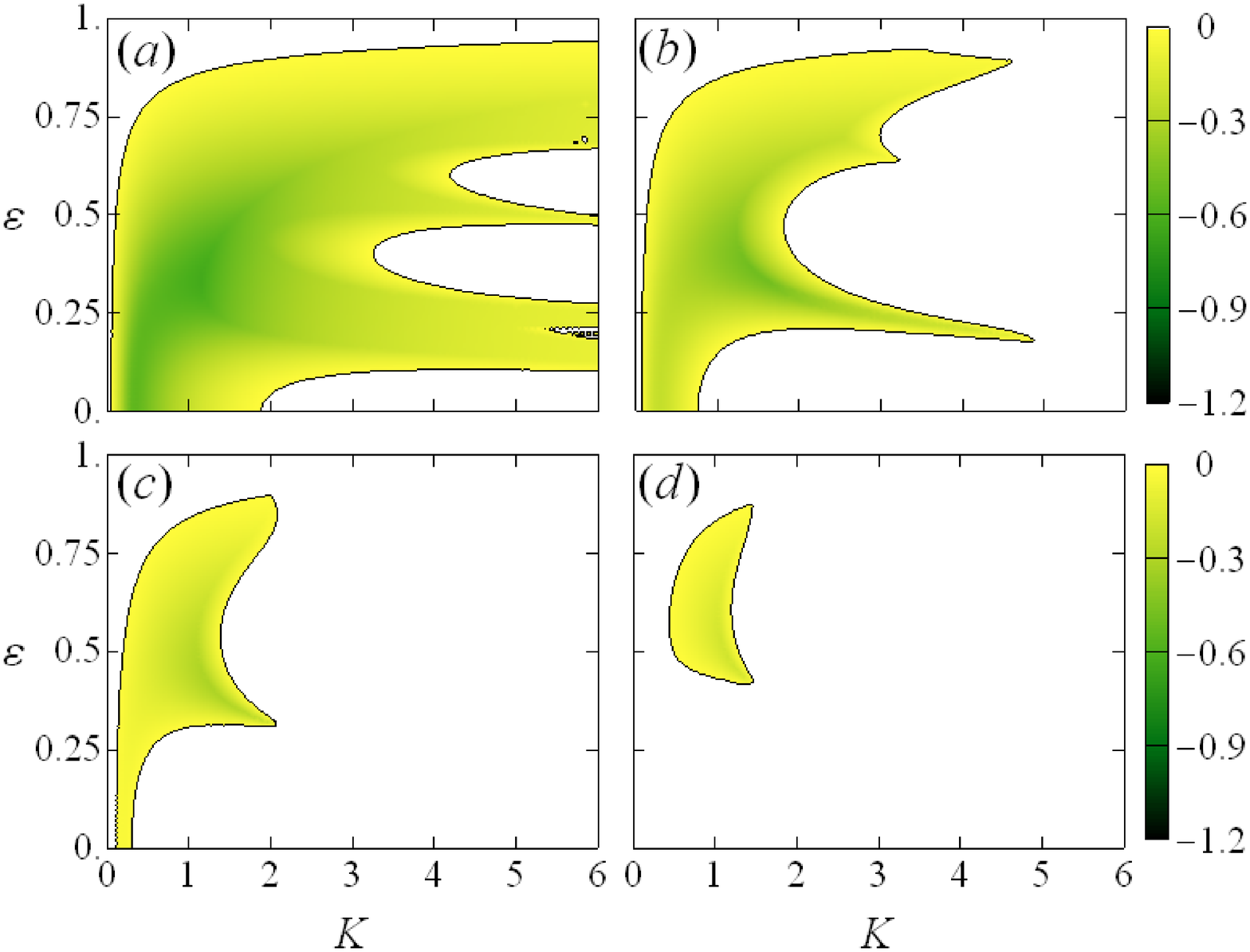}
\caption{(Color online) Control domain in the $(K,\varepsilon)$ plane for a square-wave modulation of the delay 
for different values of the latency time: (a) $\delta=0.1$, (b) $\delta=0.2$, (c) $\delta=0.3$ and (d) $\delta=0.4$. 
The other parameters are as in Fig. \ref{fig10}.}
\label{fig12}
\end{figure}

The influence of the modulation amplitude $\varepsilon$ on the stability of the fixed point can be deduced by observing the control domains in the $(K,\varepsilon)$ parameter plane.
Figures \ref{fig10}, \ref{fig11} and \ref{fig12} depict the control domains for sawtooth-, sine- and square-wave modulation, respectively, for a fixed nominal delay $\tau_0=1$ and increasing values of the latency parameter $\delta$: (a) $\delta=0.1$, (b) $\delta=0.2$, (c) $\delta=0.3$, (d) $\delta=0.4$. The parameters
of the unstable focus were set at $\lambda=0.1$ and $\omega=\pi$ as previously. It is observed that increasing the modulation amplitude 
$\varepsilon$ for a constant latency value generally leads to an enlargement of the $K$ interval of successful control, depending on the type of the
modulation. For a sufficiently large latency time when the Pyragas control fails (e.g, $\delta=0.4$ in panel (d) of Figs. \ref{fig10}-\ref{fig12}), inclusion of a variable delay in the controller makes the control possible again for a suitable choice of the modulation amplitude $\varepsilon>0$. Nevertheless, the control domain depends on the 
modulation type, and it is seen that in this case it is much less pronounced for a sawtooth-wave modulation than for the other two delay-modulation types.  


In a similar fashion, from  Eqs. (\ref{latsys1}) and (\ref{latsys2}) one obtains the parametric representation of the 
stability boundaries in the $(K,\varepsilon)$ plane
\begin{eqnarray}
K(\Omega)&=&\frac{\lambda\sin\Omega(\delta+\tau_0)+(\omega-\Omega)\cos\Omega(\delta+\tau_0)}{\sin(\Omega\tau_0)},\nonumber \\\\
\chi(i\Omega,\varepsilon)&=&\frac{\lambda\sin(\Omega\delta)+(\omega-\Omega)\cos(\Omega\delta)}{\lambda\sin\Omega(\delta+\tau_0)+(\omega-\Omega)\cos\Omega(\delta+\tau_0)},\nonumber \\
\end{eqnarray}
in which the dependence of $\varepsilon$ on the eigenfrequency $\Omega$ enters implicitly in the second equation via the function $\chi$.
The calculated boundaries of the control domain are the solid lines depicted in Figs. \ref{fig10}--\ref{fig12}.

Overall, we see a massive destruction of the control domains with increasing latency time. Compared to the effect of a phase rotation in the previous section, the presence of control-loop latency tends to spoil the control mechanism radically. Besides the destructive interference effect, on which variable-delay feedback control mainly relies, an instantaneous feedback provides the main source of stability which normally is reflected in an extensive coverage of the parameter space with solutions of successful control. If this dissipation term is replaced by a much less effective term due to latency, the control performance is consequently lost. Thus in any experimental application of variable-delay feedback control one should split the control term if possible, so that the instantaneous part is implemented by a direct modification of the system to be controlled, and the variable delay part can be realized by separate devices.

\section{Multiple delay feedback control with variable time-delays}

In order to improve the control of unstable steady states, several extensions of the Pyragas method were proposed by involving multiple delay feedback terms in the control force \cite{SSG94,SBG97,AP04,AP05}. A very efficient control scheme of this type was introduced by Ahlborn and Parlitz by utilizing a feedback force constructed from two or more independent Pyragas delayed feedback controllers with incommensurate delay times applied simultaneously in the control circuit \cite{AP04,AP05}. A key result of this multiple-delay extension was a successful stabilization of chaotic intensity fluctuations of a frequency doubled ND-doped yttrium aluminum garnet (ND:YAG) laser at higher pump rates, whose control was not achievable with a single delay controller. 

In this section, we show that this multiple delay feedback control (MDFC) can be further improved by including time-varying delays in the associated feedback terms. 
As we have shown in the previous sections, the term \textit{improvement} refers in the first instance to an extension of the parameter space of successful control. Under experimental conditions this feature is favorable if control parameters are drifting, cannot be adjusted precisely or the properties of the unstable equilibrium are unknown so that the optimal parameters of the constant delay control forces cannot be determined. Concerning the robustness of the control, we have observed for an optimal choice of modulation that the resulting maximum stability exponent in the stable domain is rather uniform. But since the complete response to perturbations of the controlled system is formed by the whole spectrum of exponents, the overall robustness can vary significantly more than reflected only by the maximum exponent. In particular, we have observed \cite{JGU12} that the excitation from the stabilized fixed point by additive noise is decreasing with increasing feedback strength in an optimal VDFC scheme although the maximum exponent does not vary much. This is another typical feature of VDFC which we regard as a major improvement compared to non-modulated methods. In the following analysis of modulated MDFC we restrict ourselves to the maximum stability exponent in order to allow for comparison with the previous sections.
For the clarity of the presentation, we consider the simplest realization of MDFC consisting of two Pyragas-type delayed feedback lines with variable time delays, but the discussion can be straightforwardly generalized to more than two delay lines. The linear normal-form system reads
\begin{widetext}
\begin{equation}
\left(
\begin{array}{c}
\dot x(t) \\
\dot y(t) \\
\end{array}
\right) =
\left(
\begin{array}{cc}
\lambda & \omega \\
-\omega & \lambda \\
\end{array}
\right)
\left(
\begin{array}{c}
x(t) \\
y(t) \\
\end{array}
\right)+K_1
\left(
\begin{array}{c}
x(t-\tau_1(t))-x(t) \\
y(t-\tau_1(t))-y(t) \\
\end{array}
\right)
+K_2
\left(
\begin{array}{c}
x(t-\tau_2(t))-x(t) \\
y(t-\tau_2(t))-y(t) \\
\end{array}
\right),
\label{parsys}
\end{equation}
where $\tau_1(t)$ and $\tau_2(t)$ are two different, time-dependent, $2\pi$-periodic delay functions:
\begin{eqnarray}
\tau_1(t)=\tau_{01}+\varepsilon_1 f_1(\nu_1 t),
\label{r6}\\
\tau_2(t)=\tau_{02}+\varepsilon_2 f_2(\nu_2 t).
\label{r7}
\end{eqnarray}
The nominal delays are denoted by $\tau_{01}$ and $\tau_{02}$, $\varepsilon_1$ and $\varepsilon_2$ are the corresponding modulation amplitudes, $\nu_1$ and $\nu_2$ are the modulation frequencies, and $f_1$ and $f_2$ are the modulation functions.
For high-frequency modulations of the delays, the system is in the distributed delay regime:
\begin{equation}
\left(
\begin{array}{c}
\dot x(t) \\
\dot y(t) \\
\end{array}
\right) =
\left(
\begin{array}{cc}
\lambda & \omega \\
-\omega & \lambda \\
\end{array}
\right)
\left(
\begin{array}{c}
x(t) \\
y(t) \\
\end{array}
\right)+K_1
\left(
\begin{array}{c}
\int_0^\infty \rho_1(\theta)x(t-\theta)\;d\theta-x(t) \\
\int_0^\infty \rho_1(\theta)y(t-\theta)\;d\theta-y(t) \\
\end{array}
\right)
+K_2
\left(
\begin{array}{c}
\int_0^\infty \rho_2(\theta)x(t-\theta)\;d\theta-x(t) \\
\int_0^\infty \rho_2(\theta)y(t-\theta)\;d\theta-y(t) \\
\end{array}
\right).
\label{pardd}
\end{equation}
\end{widetext}
Using the exponential ansatz, we obtain the characteristic equation
\begin{align}
\left[\Lambda-\lambda+K_1 \left(1-e^{-\Lambda\tau_{01}}\chi_1(\Lambda,\varepsilon_1)\right)\right.&\nonumber\\
\left.+K_2 \left(1-e^{-\Lambda\tau_{02}}\chi_2(\Lambda,\varepsilon_2)\right)\right]^2&+\omega^2=0,
\label{parcharfull}
\end{align}
where 
\begin{align}
\chi_1(\Lambda,\varepsilon_1)&=\int_{-\varepsilon_1}^{\varepsilon_1}\rho_1(\tau_{01}+\theta)e^{-\Lambda\theta}\,d\theta,\nonumber\\
\chi_2(\Lambda,\varepsilon_2)&=\int_{-\varepsilon_2}^{\varepsilon_2}\rho_2(\tau_{02}+\theta)e^{-\Lambda\theta}\,d\theta
\end{align}
are complex functions corresponding to different modulation types.
Equation (\ref{parcharfull}) can be re-cast in the simple form
\begin{align}
\Lambda+&K_1 \left(1-e^{-\Lambda\tau_{01}}\chi_1(\Lambda,\varepsilon_1)\right)\nonumber\\
+&K_2 \left(1-e^{-\Lambda\tau_{02}}\chi_2(\Lambda,\varepsilon_2)\right)=\lambda\pm i\omega.
\label{parchar}
\end{align}
The parametric representation of the stability boundary is obtained by looking for solutions of Eq. (\ref{parchar}) in the form $\Lambda=i \Omega$ and separating real and imaginary parts, 
\begin{align}
K_1&\left[1-\cos(\Omega\tau_{01})\chi_1(i\Omega,\varepsilon_1)\right]\nonumber\\
&+K_2\left[1-\cos(\Omega\tau_{02})\chi_2(i\Omega,\varepsilon_2)\right]=\lambda,
\label{parsys1}\\
K_1&\sin(\Omega\tau_{01})\chi_1(i\Omega,\varepsilon_1)\nonumber\\
&+K_2\sin(\Omega\tau_{02})\chi_2(i\Omega,\varepsilon_2)=\pm\omega-\Omega.
\label{parsys2}
\end{align}
The control parameter space is now six-dimensional $(K_1,K_2,\tau_{01},\tau_{02},\varepsilon_1,\varepsilon_2)$, but for experimental purposes it could be reduced by matching similar types of control parameters (e.g. $K_1=K_2$, or $\varepsilon_1=\varepsilon_2$). 

To gain insight into the domain structure of the multiple variable-delay feedback control, we numerically analyze Eq. (\ref{parchar}) in the parametric plane spanned by the two nominal delays $\tau_{01}$ and $\tau_{02}$. 
\begin{figure*}
\includegraphics[width=0.9\textwidth,height=!]{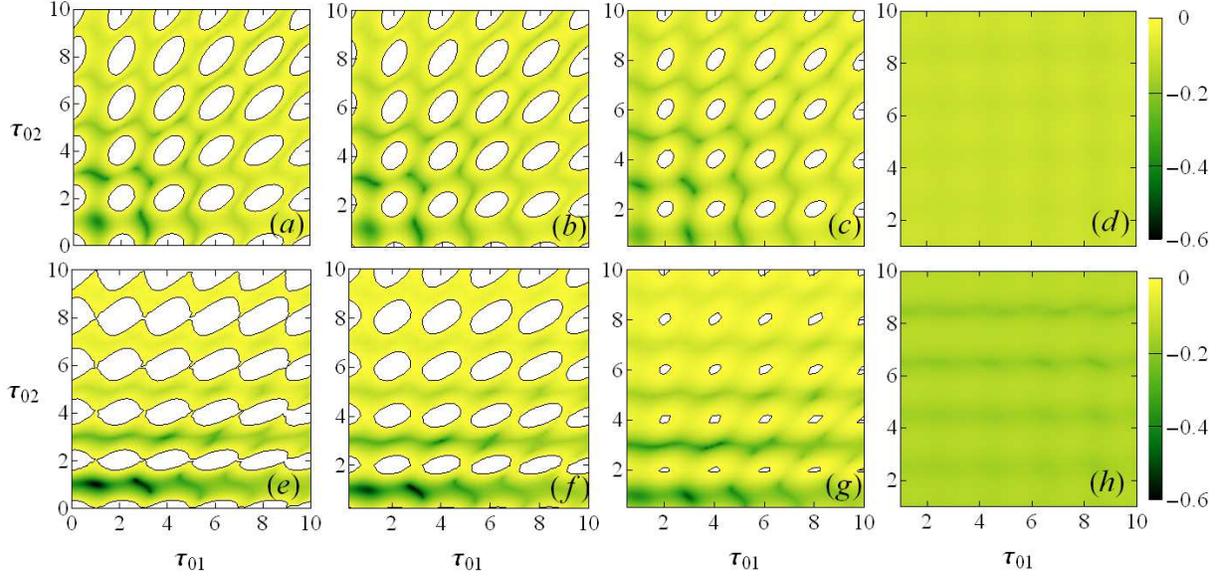}
\caption{(Color online) Control domain in the $(\tau_{01},\tau_{02})$ plane for multiple-delay feedback control (MDFC) of unstable focus with two delay lines and time-varying delays. The delays $\tau_1(t)$ and $\tau_2(t)$ in both lines are varied with sawtooth-wave modulations with matching modulation amplitudes. The feedback gain parameters are: (a)--(d)  $K_1=K_2=0.1$; (e)--(h) $K_1=0.05$ and $K_2=0.2$.  The delay modulation amplitudes are: (a,e) $\varepsilon_{1,2}=0$ (MDFC with fixed delays); (b,f) $\varepsilon_{1,2}=0.25$; (c,g) $\varepsilon_{1,2}=0.5$; (d,h) $\varepsilon_{1,2}=1$. The parameters of the unstable focus are $\lambda=0.1$ and $\omega=\pi$.
The boundary curves are given in parametric form by Eqs. (\ref{parpar1}) and (\ref{parpar2}).}
\label{fig13}
\end{figure*}
In Fig. \ref{fig13} we depict the resulting stability diagrams at different control parameter values obtained when the time delays in both feedback terms are modulated with  sawtooth-waves of equal amplitudes (i.e. $\varepsilon_1=\varepsilon_2$ at each panel). The parameters of the unstable focus are taken as $\lambda=0.1$ and $\omega=\pi$ throughout this section. The stability area is given in graytones (color online), corresponding to those values of the control parameters for which the fixed point control could be achieved. Panels (a)--(d) correspond to increasing values of $\varepsilon_{1,2}$ for a symmetrical choice of the feedback gains $K_1=K_2=0.1$. Panel (a) is the resulting stability domain without any delay variation ($\varepsilon_{1,2}=0$), i.e. MDFC with constant delays. The stability domain is symmetrical with respect to the diagonal $\tau_{01}=\tau_{02}$, and it is filled with oval-shaped instability islands (white regions) encompassing the points ($\tau_{01},\tau_{02})=(2n\pi/\omega,2m\pi/\omega)$, where $n$ and $m$ are non-negative integers. At these particular time delays, the fixed point is unstable for any values of the feedback gains $K_1$ and $K_2$. The result could be shown analytically from Eqs. (\ref{parsys1})--(\ref{parsys2}) by setting $\Omega\tau_{01}=2n\pi$ and $\Omega\tau_{02}=2m\pi$. In the same manner, we derive the conditions for the minimum feedback gains at the nominal delay values between these instability points. Specifically, at  $(\tau_{01},\tau_{02})=((2n+1)\pi/\omega,2m\pi/\omega)$ the minimum feedback gain is $K_1^\mathrm{min}=\lambda/2$; at  $(\tau_{01},\tau_{02})=(2n\pi/\omega,(2m+1)\pi/\omega)$ the minimum feedback gain is $K_2^\mathrm{min}=\lambda/2$;  at  $(\tau_{01},\tau_{02})=((2n+1)\pi/\omega,(2m+1)\pi/\omega)$ the minimum feedback gains satisfy condition $K_1^\mathrm{min}+K_2^\mathrm{min}=\lambda/2$. In the latter case, by tuning the value of one of the feedback gain parameters, a successful stabilization could be achieved at negative values of the second feedback gain parameter. 

The change in the stability diagrams as the amplitudes $\varepsilon_{1,2}$ increase from zero becomes evident from Fig. \ref{fig13}. 
Panels (b)--(d) show a monotonic increase of the control domain area for increasing modulation amplitudes: (b) $\varepsilon_{1,2}=0.25$, (c) $\varepsilon_{1,2}=0.5$, (d) $\varepsilon_{1,2}=1$. As a consequence, the instability islands become smaller, and eventually disappear at higher values of $\varepsilon_{1,2}$ (panel d). This is confirmed analytically by deriving the minimum feedback gains in the presence of delay variations when nominal delay values are integer multiples of $\pi/\omega$. By the same arguments as in the previous paragraph, we obtain: 
\begin{equation}
K_1^\mathrm{min}\left[1-\chi_1(i\omega,\varepsilon_1)\right]+K_2^\mathrm{min}\left[1-\chi_2(i\omega,\varepsilon_2)\right]=\lambda
\label{con1}
\end{equation}
at $(\tau_{01},\tau_{02})=(2n\pi/\omega,2m\pi/\omega)$,
\begin{equation}
K_1^\mathrm{min}\left[1+\chi_1(i\omega,\varepsilon_1)\right]+K_2^\mathrm{min}\left[1-\chi_2(i\omega,\varepsilon_2)\right]=\lambda
\label{con2}
\end{equation}
at $(\tau_{01},\tau_{02})=((2n+1)\pi/\omega,2m\pi/\omega)$,
\begin{equation}
K_1^\mathrm{min}\left[1-\chi_1(i\omega,\varepsilon_1)\right]+K_2^\mathrm{min}\left[1+\chi_2(i\omega,\varepsilon_2)\right]=\lambda
\label{con3}
\end{equation}
at $(\tau_{01},\tau_{02})=(2n\pi/\omega,(2m+1)\pi/\omega)$, and
\begin{equation}
K_1^\mathrm{min}\left[1+\chi_1(i\omega,\varepsilon_1)\right]+K_2^\mathrm{min}\left[1+\chi_2(i\omega,\varepsilon_2)\right]=\lambda
\label{con4}
\end{equation}
at $(\tau_{01},\tau_{02})=((2n+1)\pi/\omega,(2m+1)\pi/\omega)$.
It is seen that while the multiple delay feedback control with constant delays was unsuccessful at $(\tau_{01},\tau_{02})=(2n\pi/\omega,2m\pi/\omega)$ for any values of the feedback gain parameters $K_{1,2}$, inclusion of variable delays lifts this restriction, making the control possible with minimum gain parameters given by Eq. (\ref{con1}). 

In panels (e)--(h) of Fig. \ref{fig13} we show the corresponding increase of the control domain area for $K_1=0.05$ and $K_2=0.2$. This asymmetric choice of the feedback gains results in asymmetry of the stability domain with respect to the diagonal $\tau_{01}=\tau_{02}$, leading to horizontal enlargement of the instability islands in the absence of delay modulation (MDFC, panel a), which eventually connect into horizontal instability stripes at $\tau_{02}=2m\pi/\omega$ for lower values of the gain parameter $K_1$. By including variable delays, the stability domain increases monotonically with increasing $\varepsilon_{1,2}$, and the instability islands gradually shrink, eventually disappearing at large $\varepsilon_{1,2}$. 

\begin{figure}
\includegraphics[width=\columnwidth,height=!]{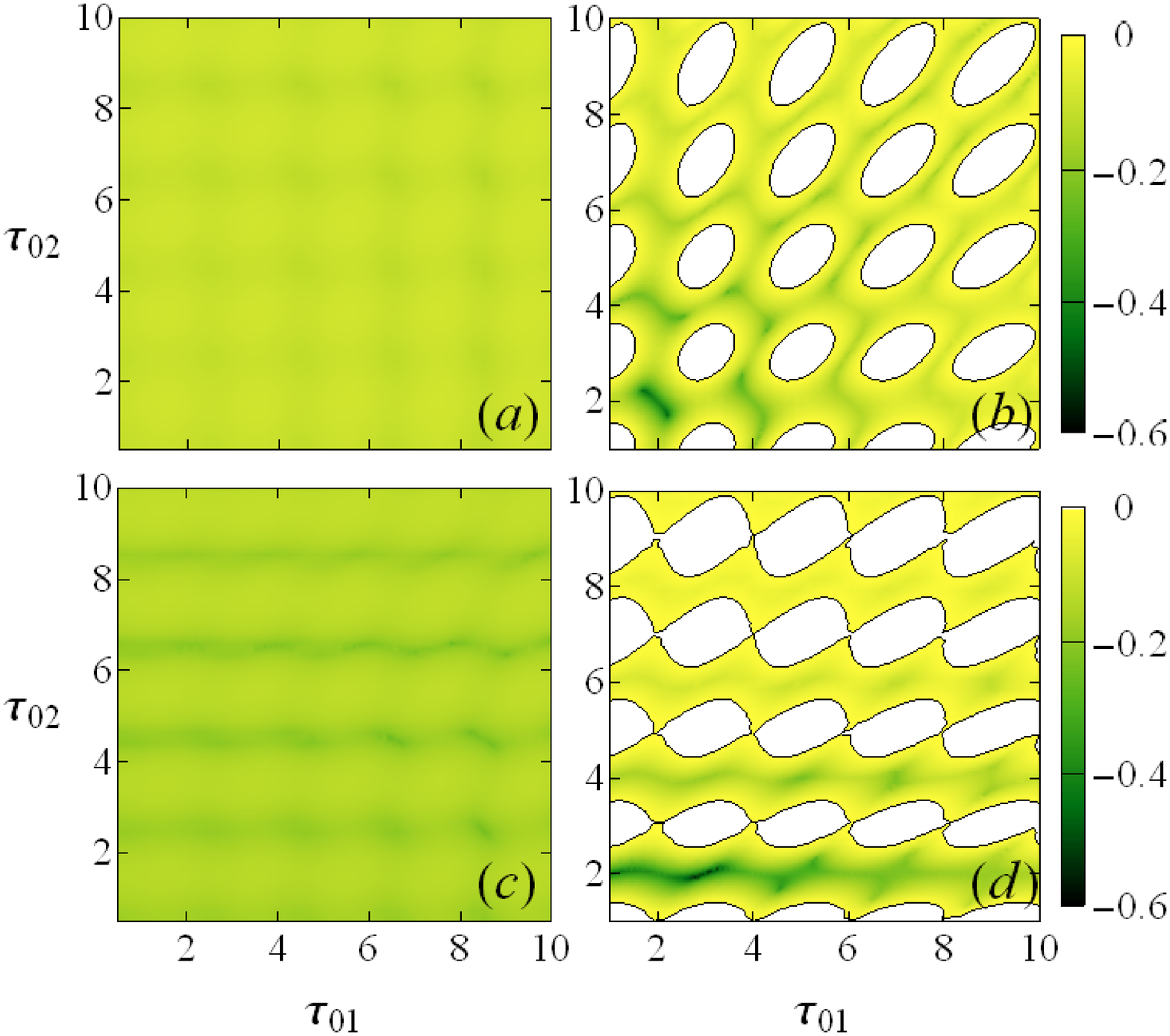}
\caption{(Color online) Control domain in the $(\tau_{01},\tau_{02})$ plane for square-wave delay modulations in both delay lines at same amplitudes. The feedback gain parameters: (a)--(b)  $K_1=K_2=0.1$; (c)--(d) $K_1=0.05$ and $K_2=0.2$.  The delay modulation amplitudes: (a,c) $\varepsilon_{1,2}=0.5$; (b,d) $\varepsilon_{1,2}=1$. The other parameters are as in Fig. \ref{fig13}.}
\label{fig14}
\end{figure}

A similar monotonic increase of the stability regions with increasing modulation amplitudes in both symmetric and asymmetric choice of the feedback gain parameters is also  observed for sine-wave modulations of the delays, and the results are not shown for compactness of the presentation. For a square-wave modulation (Fig. \ref{fig14}), we observe a non-monotonic behavior of the stability area, increasing faster with $\varepsilon_{1,2}$, attaining its maximum at $\varepsilon_{1,2}=0.5$ (panels a and c),  and then decreasing again to the same domain structure as in the unmodulated case ($\varepsilon_{1,2}=1$, panels b and d), but now with the instability islands centered at the nominal delay values being odd multiples of $\pi/\omega$ (compare with panels a and e in Fig. \ref{fig13}). This oscillatory switching of the instability islands  between $(\tau_{01},\tau_{02})=(2n\pi/\omega,2m\pi/\omega)$ and $(\tau_{01},\tau_{02})=((2n+1)\pi/\omega,(2m+1)\pi/\omega)$ continues as the modulation amplitudes $\varepsilon_{1,2}$ are further increased, and it is confirmed analytically by Eqs. (\ref{con1})--(\ref{con4}).   

\begin{figure}
\includegraphics[width=\columnwidth,height=!]{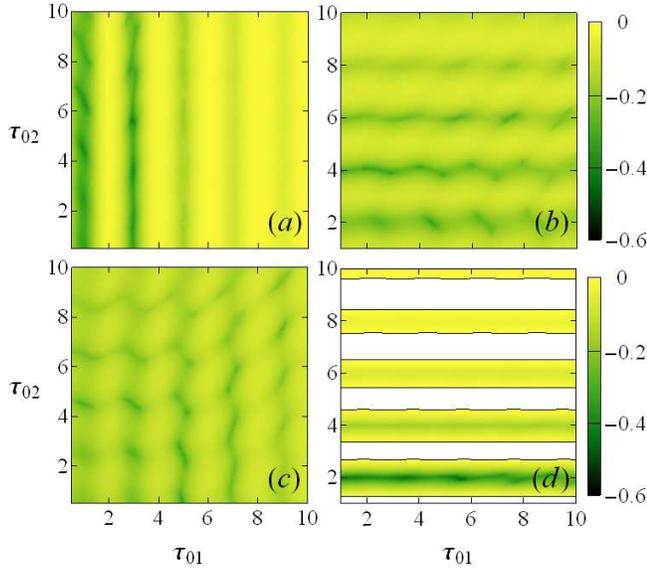}
\caption{(Color online) Control domain in the $(\tau_{01},\tau_{02})$ plane. The delay $\tau_1(t)$ is modulated with a sawtooth-wave, and $\tau_2(t)$ with a square-wave, both with the same modulation amplitude. The feedback gain parameters: (a)--(b)  $K_1=0.2$ and $K_2=0.05$; (c)--(d) $K_1=0.05$ and $K_2=0.2$.  The delay modulation amplitudes: (a,c) $\varepsilon_{1,2}=0.5$; (b,d) $\varepsilon_{1,2}=1$. The other parameters are as in Fig. \ref{fig13}.}
\label{fig15}
\end{figure}

It is also instructive to investigate the stability diagrams when the corresponding delays in the two feedback control terms are modulated with different modulation types. A general conclusion can be extracted from the sample of the results provided in Fig. \ref{fig15}. The diagrams are calculated for a sawtooth-wave variation of $\tau_1(t)$ and a square-wave variation of $\tau_2(t)$ at equal amplitudes ($\varepsilon_1=\varepsilon_2$) and asymmetric gains:  $K_1=0.2$ and $K_2=0.05$ in panels (a)--(b), and $K_1=0.05$ and $K_2=0.2$ in panels (c)--(d).  Panels (a) and (c) correspond to $\varepsilon_{1,2}=0.5$ and panels (b) and (d) to $\varepsilon_{1,2}=1$. 
The overall behavior of the control domain with increasing  $\varepsilon_{1,2}$ is mainly determined by the dominant modulation type, i.e., the modulation of the feedback with higher gain, and it is more pronounced as the difference between the respective feedback gain values becomes larger. By gradually increasing the modulation amplitudes, the domain is enlarged monotonically in panels (a)--(b) where the feedback term with a sawtooth-wave modulation dominates over the square-wave term ($K_1>K_2$). The reversed case scenario ($K_2>K_1$) in panels (c)--(d) shows a non-monotonic behavior typical for a square-wave modulation by the appearance of instability stripes oscillating between the values of $\tau_{02}$ being even and odd multiples of $\pi/\omega$.

To derive the equations for the boundary curves in Figs. (\ref{fig13})--(\ref{fig15}), we follow the approach in Ref. \cite{GNC05} and rewrite the characteristic Eq. (\ref{parchar}) as
\begin{equation}
1+a(\Lambda)e^{-\Lambda\tau_{01}}+b(\Lambda)e^{-\Lambda\tau_{02}}=0,
\label{niculescueq}
\end{equation}
where $a(\Lambda)$ and $b(\Lambda)$ are given by
\begin{align}
a(\Lambda)&=-\frac{K_1\chi_1(\Lambda,\varepsilon_1)}{\Lambda-\lambda\mp i\omega+K_1+K_2},
\label{alambda}\\
b(\Lambda)&=-\frac{K_2\chi_2(\Lambda,\varepsilon_2)}{\Lambda-\lambda\mp i\omega+K_1+K_2}.
\label{blambda}
\end{align}
At the control boundary ($\Lambda=i\Omega$) the three terms in Eq. (\ref{niculescueq}) can be considered as three vectors in the complex plane, with the corresponding magnitudes 1, $|a(i\Omega)|$ and $|b(i\Omega)|$. According to Eq. (\ref{niculescueq}), the sum of these vectors is a zero vector, and from the triangle formed by the vectors it is straightforward to obtain the parametric representation of $\tau_{01}$ and $\tau_{02}$ on the eigenfrequency $\Omega$:
\begin{align}
\tau_{01}(\Omega)&=\frac{\mathrm{Arg}\left[a(i\Omega)\right]+(2u-1)\pi\pm\alpha_1}{\Omega}\geq 0,\nonumber\\
&u=u_0^\pm,u_0^\pm+1,u_0^\pm+2\dots,
\label{parpar1}\\ \nonumber\\
\tau_{02}(\Omega)&=\frac{\mathrm{Arg}\left[b(i\Omega\right]+(2v-1)\pi\mp\alpha_2}{\Omega}\geq 0,\nonumber\\
&v=v_0^\pm,v_0^\pm+1,v_0^\pm+2\dots,
\label{parpar2}
\end{align}
where $u_0^\pm$ and $v_0^\pm$ are the smallest possible integers such that the corresponding values of $\tau_{01}$ and $\tau_{02}$ are all non-negative, and $\alpha_1,\alpha_2\in[0,\pi]$ are the internal angles of the triangle formed by the vectors, calculated from the law of cosines as:
\begin{align}
\alpha_1&=\mathrm{Arccos}\left(\frac{1+|a(i\Omega)|^2-|b(i\Omega)|^2}{2|a(i\Omega)|}\right),\\
\alpha_2&=\mathrm{Arccos}\left(\frac{1+|b(i\Omega)|^2-|a(i\Omega)|^2}{2|b(i\Omega)|}\right).
\end{align}  

\begin{figure}
\includegraphics[width=\columnwidth,height=!]{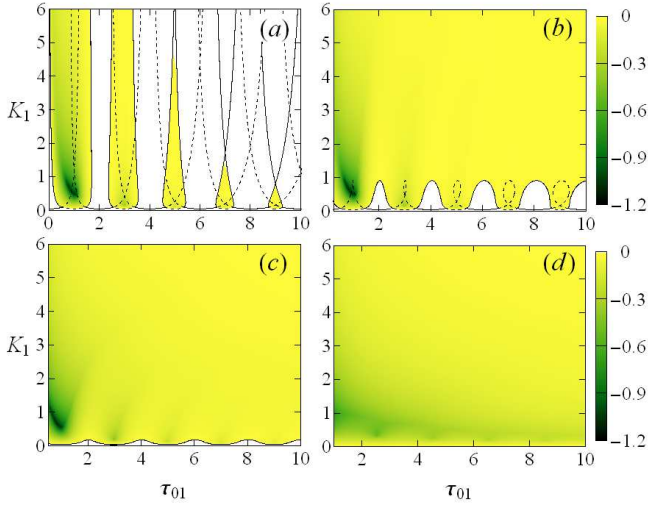}
\caption{(Color online) Control domain in the $(K_1,\tau_{01})$ plane for a sawtooth-wave modulation in both delay lines for different values of the modulation amplitudes: (a) $\varepsilon_{1,2}=0$ (MDFC, constant delays), (b) $\varepsilon_{1,2}=0.25$, (c) $\varepsilon_{1,2}=0.5$ and (d) $\varepsilon_{1,2}=1$. 
The feedback gain and the nominal delay of the second feedback line are fixed at $K_2=0.1$ and $\tau_{02}=2$. 
The solid/dashed lines that describe the stability boundary are calculated from Eqs. (\ref{park})--(\ref{partau2}).
The parameters of the free-running system are as in Fig. \ref{fig13}.}
\label{fig16}
\end{figure}

To gain further insight into the superiority of MDFC with time-varying delays with respect to the fixed delays realization, we have numerically calculated the stability diagrams in the parametric plane of the nominal delay $\tau_{01}$ and the feedback gain $K_1$ of the first feedback line at different $\varepsilon_{1,2}$, by fixing the nominal delay $\tau_{02}$ and the feedback gain $K_2$ of the second feedback line. For the value $\tau_{02}$ we choose an even multiple of $\pi/\omega$ at which the fixed point control by MDFC with constant delays always fails if $\tau_{01}$ is also an even multiple of $\pi/\omega$. In Fig. \ref{fig16} we show the corresponding control domains for $\tau_{02}=2\pi/\omega=2$ and $K_2=0.1$ for sawtooth-wave modulations of the delays at different modulation amplitudes: (a) $\varepsilon_{1,2}$=0 (MDFC), (b) $\varepsilon_{1,2}$=0.25, (c) $\varepsilon_{1,2}$=0.5, (d) $\varepsilon_{1,2}$=1. For fixed delays (MDFC case, panel a), the control domain is consisted of isolated stability islands encompassing $\tau_{01}=(2n+1)\pi/\omega$, disconnected by instability stripes at $\tau_{01}=2n\pi/\omega$ at which control fails at any $K_{1,2}$. As the modulation amplitudes increase (panels b-d), the stability islands gradually expand into a connected stability region, making the control possible at any $\tau_{01}$. This monotonic expansion of the stability domain is sustained in the case of a sine-wave modulation, but not for a square-wave modulation, as expected from the earlier analysis. In the latter case, instability stripes are disappearing and re-appearing again with increasing $\varepsilon_{1,2}$ in an oscillatory manner at even multiples of $\pi/\omega$ along $\tau_{01}$ axis. When the delays are modulated with different modulation types, the behavior of the domain structure with increasing $\varepsilon_{1,2}$ is determined by the dominant feedback term. 

The parametric representation of the stability boundaries in Fig. \ref{fig16} parametrized by the eigenfrequency $\Omega$ can be obtained from Eqs. (\ref{parsys1})--(\ref{parsys2}):
\begin{align}
K_1(\Omega)&=\frac{p\pm\sqrt{p^2\chi_1^2(i\Omega,\varepsilon_1)-\left[1-\chi_1^2(i\Omega,\varepsilon_1)\right]q^2}}{1-\chi_1^2(i\Omega,\varepsilon_1)},\label{park}\\
\tau_{01}(\Omega)&=\frac{1}{\Omega}\left[\arcsin\left(\frac{q}{K_1(\Omega)\chi_1(i\Omega,\varepsilon_1)}\right)+2n\pi\right],\label{partau1}\\
\tau_{02}(\Omega)&=\frac{1}{\Omega}\left[-\arcsin\left(\frac{q}{K_1(\Omega)\chi_1(i\Omega,\varepsilon_1)}\right)+(2n+1)\pi\right],
\label{partau2}
\end{align}
where for brevity we used the notations
\begin{align}
p&=\lambda-K_2\left[1-\cos(\Omega\tau_{02})\chi_2(i\Omega,\varepsilon_2)\right],\\
q&=\pm\omega-\Omega-K_2\sin(\Omega\tau_{02})\chi_2(i\Omega,\varepsilon_2).
\end{align}

Finally, one might ask the question, to which extent the extended time-delayed feedback control (ETDFC) by Socolar et al. \cite{SSG94,SBG97} is also affected by a modulation on top of the discrete exponential delay distribution created by the original method. In Fig. \ref{fig13} we have already presented the results for an asymmetric choice of the coupling gains $K_1=0.05$ and $K_2=0.2$. 
With a corresponding choice of $\tau_{01} = 2\tau_{02}$, which gives a special section through the shown parameter planes, the control scheme can already be seen as a rudimentary ETDFC scheme with $\tau_0 = \tau_{02}$, in which the longer delays $n \tau_0$ with $n>2$ are neglected.
A fully developed ETDFC scheme together with a modulation of the delays would not show any important features that are not already documented in the modulated MDFC.

\section{Conclusions}

Extension of the delayed feedback control of unstable equilibria by introducing time-varying delays in one or more independent delayed feedback lines in the control circuit significantly enlarges the stability domain in the control parameter space. In addition to the domain enlargement, the variable-delay feedback method is successful in stabilizing unstable states even for those cases where the control always fails in the constant delay case.

We have shown that an analytical investigation of the control domains becomes possible in the range of high-frequency delay modulations, in which case the variable-delay term in the controller can be approximated by an equivalent distributed delay term. Our approach is motivated by both simulations and real experiments \cite{JGU12}, which suggest that when the frequency of the delay modulation is comparable to the system frequencies, the system dynamics can be very well approximated with the distributed-delay effect.
Restricting our attention to this high-frequency limit, we have investigated the control efficiency of the variable-delay feedback method by considering a simple 
normal form of an unstable focus and two different realizations of the controller which are experimentally relevant: a non-diagonal coupling of the control force 
realized via a phase-dependent coupling matrix, and the incorporation of an additional small constant delay term in all the arguments of the feedback force that represents 
the latency of the control-line realization. In addition, we have explored a simple realization of a multiple delay feedback controller consisting of two independent delay lines of Pyragas type with time-varying delays.
In each case, the variable-delay feedback control with a finite modulation of the delay is shown superior with respect to the constant delay case. This renders the proposed control method promising for further practical implementations in real experiments.

The two-dimensional linear system used in the analysis is generic for all systems with unstable fixed points of a focus type, preserving the essential features of the higher-dimensional dynamics around the equilibrium. Thus, the results obtained in this paper aim to give a succinct explanation of the delayed feedback control mechanism with time-dependent delay in real systems (regular or chaotic) with similar bifurcation properties.
As in the original time-delayed feedback control, the torsion of the orbit is a necessary condition for the control method to be able to stabilize equilibria.

Since in practice the equivalence of the distributed delay case holds already for fairly low modulation frequencies, we propose that variable-delay feedback is a convenient experimental method for realizing distributed delay feedback with different delay distribution.

\begin{acknowledgments}
This work was supported by Deutsche Forschungsgemeinschaft in the framework of SFB 910: ``Control of self-organizing
nonlinear systems: Theoretical methods and concepts of application.'' 
Thomas J\"ungling acknowledges support by FEDER (EU) under the project FISICOS (FIS2007-60327).
\end{acknowledgments}

\end{document}